\documentclass[preprint2,times,tighten,twocolappendix]{aastex6}
\pdfoutput=1 
\usepackage{amsmath,amstext}
\usepackage[figure,figure*]{hypcap}
\usepackage{newtxmath} 
\usepackage{multirow}

\hypersetup{linkcolor=blue,citecolor=blue,filecolor=cyan,urlcolor=magenta}

\def \note #1 {\texttt{NOTE: #1}}
\def \tbd #1 {{\textcolor{red}{! #1 !} }}
\def \bo #1 {{\bf #1\ }}

\def \suzaku {{\em Suzaku}}
\def \nustar {{\em NuSTAR}}

\def \swiftbat {{\em Swift}/BAT}

\def \hitomi {{\em Hitomi}}
\def \chandra {{\em Chandra}}
\def \mytorus {\texttt{MYtorus}}
\def \bntorus {\texttt{BNtorus}}

\def \ctorus {\texttt{ctorus}}
\def \etorus {\texttt{etorus}}
\def \borus {\texttt{BORUS}}
\def \borustwo {\texttt{borus02}}
\def \pexrav {\texttt{pexrav}}
\def \pexmon {\texttt{pexmon}}

\def \xspec {\texttt{Xspec}}

\def \nh {$N_{\mbox{\scriptsize H}}$}

\def \lognhtor {$\log\,N_{\mbox{\scriptsize H,tor}}$}
\def \lognhtorcmmt {$\log\,N_{\mbox{\scriptsize H,tor}}/\mbox{cm}^{-2}$}
\def \nhtor {$N_{\mbox{\scriptsize H,tor}}$}

\def \lognhloscmmt {$\log\,N_{\mbox{\scriptsize H,los}}/\mbox{cm}^{-2}$}

\def \nhlos {$N_{\mbox{\scriptsize H,los}}$}

\def \rpex {$R_{\mbox{\scriptsize pex}}$}
\def \cftor {$C_{\mbox{\scriptsize tor}}$}
\def \thetator {$\theta_{\mbox{\scriptsize tor}}$}
\def \thetainc {$\theta_{\mbox{\scriptsize inc}}$}
\def \cosi {$\cos\,\theta_{\mbox{\scriptsize inc}}$}
\def \ecut {$E_{\mbox{\scriptsize cut}}$}
\def \pnull {$p_{\mbox{\scriptsize null}}$}

\def \feew {{EW$_{\mbox{\scriptsize Fe\,K}\alpha}$}}

\def \afe {{$A_{\mbox{\scriptsize Fe}}$}}
\def \afesun {{$A_{\mbox{\scriptsize Fe,}\,\odot}$}}
\def \cmmt {cm$^{-2}$}

\shorttitle{Torus Covering Factors from Broadband X-ray Spectroscopy}
\shortauthors{Balokovi\'{c} et al.}

\begin{document}

\title{New Spectral Model for Constraining Torus Covering Factors\\from Broadband X-ray Spectra of Active Galactic Nuclei}

\author{M. Balokovi\'{c}\altaffilmark{1,2}, M. Brightman\altaffilmark{1}, F.\,A. Harrison\altaffilmark{1}, A. Comastri\altaffilmark{3},\\ C. Ricci\altaffilmark{4,5,6}, J. Buchner\altaffilmark{4,7}, P. Gandhi\altaffilmark{8}, D. Farrah\altaffilmark{9}, D. Stern\altaffilmark{10}}

\begin{center}
\altaffiltext{1}{Cahill Center for Astronomy and Astrophysics, California Institute of Technology, Pasadena, CA 91125, USA}
\altaffiltext{2}{Harvard-Smithsonian Center for Astrophysics, 60 Garden Street, Cambridge, MA 02140, USA}
\altaffiltext{3}{INAF -- Osservatorio Astronomico di Bologna, via Gobetti 93/3, 40129 Bologna, Italy} 
\altaffiltext{4}{Instituto de Astrof\'{i}sica, Facultad de F\'{i}sica, Pontificia Universidad Cat\'{o}lica de Chile, Casilla 306, Santiago 22, Chile}
\altaffiltext{5}{Chinese Academy of Sciences South America Center for Astronomy and China-Chile Joint Center for Astronomy, Camino El Observatorio 1515, Las Condes, Santiago, Chile}
\altaffiltext{6}{Kavli Institute for Astronomy and Astrophysics, Peking University, China}
\altaffiltext{7}{Millenium Institute of Astrophysics, Vicu\~{n}a MacKenna 4860, 7820436 Macul, Santiago, Chile}
\altaffiltext{8}{Department of Physics and Astronomy, University of Southampton, Highfield, Southampton SO17 1BJ, UK}
\altaffiltext{9}{Department of Physics, Virginia Tech, Blacksburg, VA 24061, USA}
\altaffiltext{10}{Jet Propulsion Laboratory, California Institute of Technology, 4800 Oak Grove Drive, Pasadena, CA 91109, USA}
\end{center}

\begin{abstract}
The basic unified model of active galactic nuclei (AGN) invokes an anisotropic obscuring structure, usually referred to as a torus, to explain AGN obscuration as an angle-dependent effect. We present a new grid of X-ray spectral templates based on radiative transfer calculations in neutral gas in an approximately toroidal geometry, appropriate for CCD-resolution X-ray spectra (FWHM$\geqslant$130\,eV). Fitting the templates to broadband X-ray spectra of AGN provides constraints on two important geometrical parameters of the gas distribution around the supermassive black hole: the average column density and the covering factor. Compared to the currently available spectral templates, our model is more flexible, and capable of providing constraints on the main torus parameters in a wider range of AGN. We demonstrate the application of this model using hard X-ray spectra from \nustar\ (3--79\,keV) for four AGN covering a variety of classifications: 3C\,390.3, NGC\,2110, IC\,5063 and NGC\,7582. This small set of examples was chosen to illustrate the range of possible torus configurations, from disk-like to sphere-like geometries with column densities below, as well as above, the Compton-thick threshold. This diversity of torus properties challenges the simple assumption of a standard geometrically and optically thick toroidal structure commonly invoked in the basic form of the unified model of AGN. Finding broad consistency between the our constraints and those from infrared modeling, we discuss how the approach from the X-ray band complements similar measurements of AGN structures at other wavelengths.
\end{abstract}

\maketitle

\section{Introduction}
\label{sec:intro}

According to the simple unification model of active galactic nuclei (AGN), a toroid-like structure (popularly, {\em torus}) provides the anisotropic obscuration needed to explain the diversity of AGN observed across the electromagnetic spectrum (\citealt{antonucci-1993}, \citealt{urry+padovani-1995}). The torus absorbs and reprocesses radiation from the accretion disk and the innermost regions around the supermassive black hole (SMBH). Reprocessed thermal emission from dust in the torus is observed primarily in the infrared part of the spectrum (see, e.g., \citealt{hoenig-2012}, \citealt{netzer-2015} for recent reviews). Signatures of reprocessing in the X-ray band---narrow fluorescent emission lines (most notably, neutral iron lines around 6.4\,keV) and the Compton hump broadly peaking at 10--30\,keV---arrise primarily from interaction of X-ray photons with the surrounding gas (e.g., \citealt{leahy+creighton-1993}, \citealt{ghisellini+1994}, \citealt{krolik+1994}). These spectral features have been observed in nearly all X-ray spectra of non-blazar AGN with sufficient energy coverage and data quality (e.g., \citealt{nandra+pounds-1994}, \citealt{turner+1997}, \citealt{risaliti-2002}, \citealt{dadina-2008}, \citealt{rivers+2013}, \citealt{vasudevan+2013}, \citealt{kawamuro+2016}).

A large body of literature on X-ray spectroscopy of AGN is based on models computed for reprocessing in a semi-infinite plane geometry, the most popular of which is \pexrav\ \citep{magdziarz+zdziarski-1995}. Spectral models in which the signature of the torus is approximated with \pexrav\ have been popular for describing the phenomenology of broadband X-ray spectra of AGN because this simple geometry is easily parametrized, and because the quality of hard X-ray data ($>10$\,keV) was such that deviations from this assumption were generally not considered significant. Reprocessed continua are known to vary as a function of geometry of the reprocessing material (e.g., \citealt{nandra+george-1994}, \citealt{yaqoob-1997}, \citealt{murphy+yaqoob-2009}, \citealt{ikeda+2009}, \citealt{brightman+nandra-2011a}, \citealt{liu+li-2014}, \citealt{furui+2016}); however, the ability to constrain the geometry of the reprocessing material is clearly lacking in the \pexrav-based phenomenological approach. The 100-fold increase in sensitivity in the hard X-ray band ($>10$\,keV) brought by \nustar\ \citep{harrison+2013} made it possible to study the spectral signatures of the torus in detail.

Empirically motivated spectral models with approximately toroidal geometry have been calculated by \citet[][\mytorus\ hereafter]{murphy+yaqoob-2009}, \citet{ikeda+2009}, \citet[][\bntorus\ hereafter]{brightman+nandra-2011a}, \citet[][{\ctorus} hereafter]{liu+li-2014} and \citet{furui+2016}, and some were made available to the community. These models, especially \mytorus\ and \bntorus, have been used extensively for detailed spectroscopic studies of nearby obscured AGN observed with \nustar\ (e.g., \citealt{arevalo+2014-circinus}, \citealt{balokovic+2014}, \citealt{annuar+2015-ngc5643}, \citealt{rivers+2015-ngc7582}, \citealt{ricci+2016-ic751}, \citealt{boorman+2016-ic3639}, \citealt{gandhi+2017-ngc7674}), as well as studies using broadband data with hard X-ray coverage from {\em Suzaku}/PIN, {\em Swift}/BAT and {\em INTEGRAL} instruments (e.g., \citealt{fukazawa+2011}, \citealt{tazaki+2011}, \citealt{yaqoob-2012-ngc4945}, \citealt{braito+2013-ngc4507}, \citealt{vasylenko+2013-ngc3281}, \citealt{miniutti+2014-eso323}, \citealt{yaqoob+2015-mrk3}).

These torus models are limited in the range of physical scenarios they describe. In \mytorus, with a geometry of an actual torus, the covering factor is fixed (50\,\% of the sky covered as seen from the SMBH). This assumption limits the range of spectral shapes that the model can reproduce without {\em decoupling} it into several independent components. This leads to the normalizations of the reprocessed spectrum no longer being consistent with the torus geometry. Such a decoupling is often required in spectral analyses of high-quality broadband X-ray spectra (e.g., \citealt{puccetti+2014-ngc4945}, \citealt{bauer+2015-ngc1068}, \citealt{guainazzi+2016-mrk3}). As described in detail by \citet{yaqoob-2012-ngc4945}, the user typically needs to assume presence of reprocessing spectral components for both edge-on and pole-on inclination with arbitrary relative normalization, disconnected from the normalization of the observed intrinsic continuum.

In the \bntorus\ model, the torus opening angle is a free parameter, but the torus column density is assumed to be equal to the line-of-sight column density (\nhlos) for any obscured AGN. While this assumption does not hold in general (e.g., \citealt{risaliti+2010-mrk1210}, \citealt{marchese+2012-ngc454}, \citealt{yaqoob+2015-mrk3}), and is also dependent on the specific modeling used (i.e., phenomenological versus physically motivated), it does describe some AGN well (e.g., \citealt{gandhi+2014-mrk34}, \citealt{annuar+2015-ngc5643}, \citealt{koss+2015-ngc3393}). In particular, Compton-thick (CT; \nh\,$>10^{24}$\,\cmmt) AGN represent 20--50\,\% of the local AGN population (e.g., \citealt{ricci+2015}, \citealt{akylas+2016}, \citealt{koss+2016}), and it is widely believed that our line of sight crosses their tori in most cases \citep{ricci+2017-nature}. \bntorus\ may therefore be applicable to CT AGN spectra, and \citet{brightman+2015} used it to measure the torus covering factors in a sample of 10~\nustar-observed CT AGN.

Many multi-epoch X-ray studies have shown that \nhlos\ varies on timescales of hours to months, as clouds of gas pass in and out of our line of sight (e.g., \citealt{risaliti+2002}, \citealt{lamer+2003-ngc3227}, \citealt{risaliti+2010-mrk1210}, \citealt{marchese+2012-ngc454}, \citealt{braito+2013-ngc4507}, \citealt{markowitz+2014}, \citealt{rivers+2015-ngc7582}, \citealt{guainazzi+2016-mrk3}, \citealt{marinucci+2016-ngc1068}, \citealt{ricci+2016-ic751}). The {\em average} column density of the torus, which is a large parsec-scale structure, can only vary over significantly longer timescales ($\gtrsim$year). The ability to decouple the line-of-sight component from reprocessing in the spatially extended torus is essential for multi-epoch modeling of a wide variety of AGN. While this is possible with \mytorus\ and \ctorus, they do not feature the covering factor as a free parameter.

The covering factor of the torus is one of its most basic geometric parameters. It may be affected by winds and outflows from the innermost regions around the SMBH, therefore providing insight into physics of AGN feedback and the interaction of SMBHs with their host galaxies (e.g., \citealt{elvis-2000}, \citealt{hopkins+2006}, \citealt{fabian-2012}, \citealt{heckman+best-2014}, \citealt{netzer-2015}). Studies in the infrared band indicate that the torus covering factor may be a function of luminosity (e.g., \citealt{maiolino+2007}, \citealt{treister+2008}, \citealt{assef+2013}), and may correlate with other measurable properties (e.g., presence of broad lines in optical spectra; \citealt{mateos+2016}). It has been suggested that the covering factor depends on the Eddington ratio (e.g., \citealt{ezhikode+2016}, \citealt{buchner+bauer-2017}, \citealt{ricci+2017-nature}), and that its dependence on luminosity or the Eddington ratio changes with redshift (e.g., \citealt{aird+2015}, \citealt{buchner+2015}). AGN population studies in the X-ray band suggest that the fraction of obscured AGN drops as a function of luminosity (e.g., \citealt{sazonov+revnivtsev-2004}, \citealt{hasinger-2008}, \citealt{burlon+2011}, \citealt{vasudevan+2013}). A tentative trend for lower covering factors at higher luminosity was also found from analyses of individual AGN both in the infrared \citep{alonsoHerrero+2011} and in the X-ray band \citep{brightman+2015}.

The next step toward systematically probing the properties of the torus from the X-ray band will be analyses of AGN samples with good-quality hard X-ray data. A large, representative sample of nearby obscured AGN observed with \nustar\ will be presented in an upcoming paper (B18 hereafter).\footnote{See \citet{balokovic-phd} for preliminary results.} This study revealed that the local AGN population exhibits a broad range of Compton hump strengths when modeled with \pexrav\footnote{In \pexrav\ and \pexmon\ (extension of \pexrav, including fluorescent line emission; \citealt{nandra+2007}), the contribution of reprocessed continuum is parametrized with the spectral parameter $R$. To avoid confusion, because this parameter can formally take on negative values, we define \rpex\,$=|R|$. \rpex\,$=1$ corresponds to the amount of reprocessing created by an infinitely optically thick plane covering one half of the sky as seen from the X-ray source. While small deviations from unity can be interpreted as the reprocessing medium covering a solid angle of $\simeq 2 \pi$\,\rpex, this interpretation clearly fails for deviations greater than a factor of $\simeq2$, which are often found in X-ray spectral analyses.}, including a significant fraction with high values (e.g., \citealt{ricci+2011}, \citealt{rivers+2013}, \citealt{vasudevan+2013}, \citealt{kawamuro+2016}, B18), which may be indicative of the increased (or decreased) prominence of the Compton hump as a function of the covering factor of the torus and its average column density. While this idea is not new (e.g., \citealt{madejski+2000-ngc4945,krolik+1994,ghisellini+1994}), the operational tool for measuring the covering factor from X-ray spectra independently from the line-of-sight component has thus far not been available.

In this paper we present a new tool for probing the torus structure from the X-ray band. With its increased flexibility in comparison with currently available models, we aim to enable studies of the main torus parameters in AGN of any class. Our grid of spectral templates is made available to the community in the form of a new \xspec\ table model \citep{arnaud-1996}. Construction of the spectral template grid is presented in \S\,\ref{sec:model}. In \S\,\ref{sec:fitting} we demonstrate its use on \nustar\ spectra of four different AGN in order to highlight its features and capabilities. In \S\,\ref{sec:discussion} we briefly discuss the results for this small and diverse set of examples, and their interpretation. We also make a comparison to relevant measurements from the literature, with particular emphasis on the infrared, and discuss the prospect for future synergy with other methods of constraining torus geometry.

\section{New Spectral Templates}
\label{sec:model}

Reprocessed components of AGN X-ray spectra are formed in interaction of the intrinsic X-ray continuum of AGN with the surrounding medium. In order to investigate the details of the complex relationship between the geometry of this material and the observed spectra, we have built a new Monte Carlo radiative transfer code \borus\ (Balokovi\'{c} et al., in prep.). Radiative transfer simulations using this code can be performed in an arbitrary geometry, and at energy resolution matching high-resolution X-ray calorimeters similar to {\em Hitomi}/SXS \citep{mitsuda+2014}, e.g., {\em Athena}/XIFU \citep{barret+2016}, and the \hitomi\ successor {\em XARM} (X-ray Astronomy Recovery Mission). Details of the radiative transfer calculations in a range of geometries appropriate for AGN tori will be presented in the aforementioned paper; here we only outline the main properties, and then focus on the particular subset of low-resolution spectral templates used in this paper. The spectral templates are available on the Web\footnote{\url{http://www.astro.caltech.edu/~mislavb/download}}, and can be obtained directly from the authors.

\subsection{Model Setup}
\label{sec:model-setup}

\begin{figure}[t]
\begin{center}
\includegraphics[width=\columnwidth]{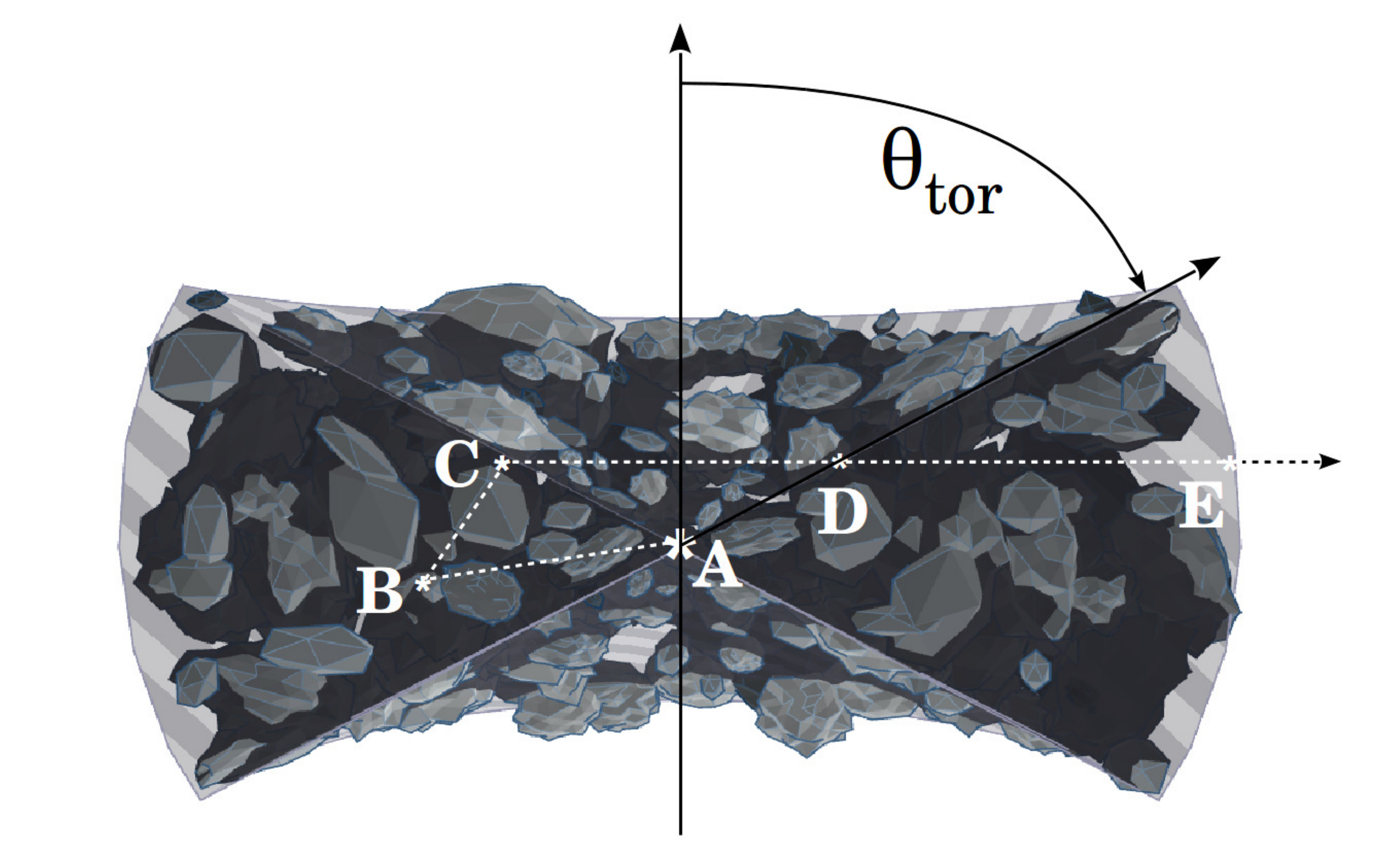}
\caption{Cross-section of the approximately toroidal geometry adopted for our model. Though the torus may be composed of individual clouds in reality (shown as light grey blobs), we approximate it with a uniform density sphere with two conical cutouts (shown as striped semi-transparent geometrical shape). The half-opening angle of the torus, \thetator\ (or, equivalently, the covering factor, \cftor\,$=\cos$\,\thetator) is a free parameter of our model. The white asterisk in the middle (point A) represents the X-ray source. White dashed lines and letters trace a particular photon ray which exits the system in the direction of the observer to the right (looking at the system edge-on). Note that in a clumpy torus it may happen that photons scattered toward the observer at point C (near the inner edge of the torus) can escape without absorption if they pass between individual clouds. However, in a torus with uniform density, such photons will undergo absorption and scattering between points D and E. The difference in resulting reprocessed spectra is illustrated in \autoref{fig:modcomp}. \label{fig:sketch} }
\end{center}
\end{figure}

\begin{figure*}
\begin{center}
\includegraphics[width=0.9\textwidth]{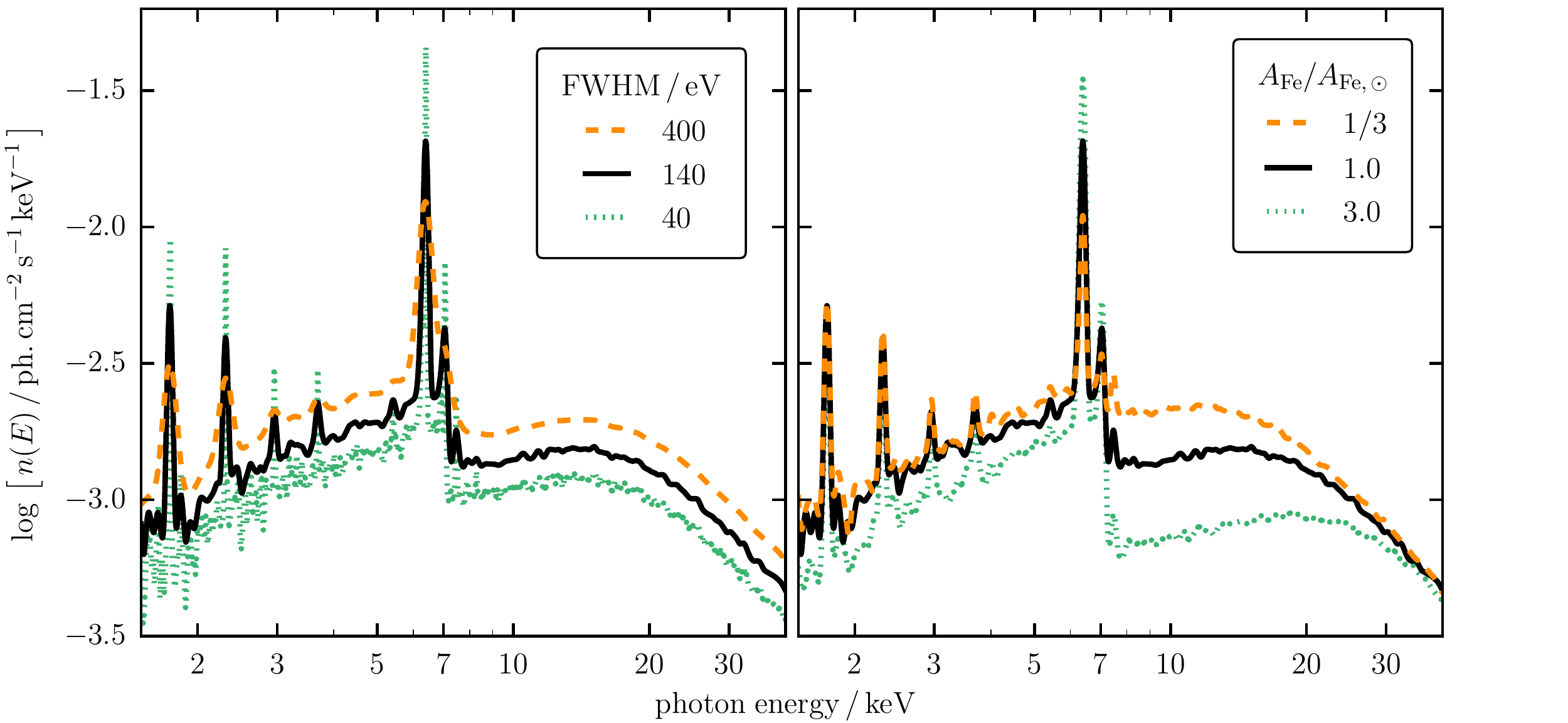}
\caption{{\em Left panel}: Comparison of \borustwo\ reprocessed spectra at different instrumental resolutions, in terms of FWHM at 6.4\,keV: 400\,eV (dashed orange line, matching \nustar\ resolution), 140\,eV (solid black line, typical for CCD-based imaging instruments), and 40\,eV (dotted green line, achievable with \chandra/HETGS). For clarity, the spectra are offset in normalization by 25\,\%. The spectra shown here are for $\Gamma=1.8$, torus column density $3\times10^{24}$\,\cmmt, 50\,\% covering factor, and viewing angle equal to the opening angle of the torus. {\em Right panel}: Spectra for different values of the relative iron abundance ($A_{\mbox{\small Fe}}$), showing significant changes in the shape of the reprocessed continuum in addition to the intensity of the Fe fluorescence lines. The black solid line shows the same spectrum in both panels. \label{fig:repcomp_extra} }
\end{center}
\end{figure*}

The \borus\ radiative transfer code is capable of computation in an arbitrary 3-dimensional space within which matter density can be represented as a mathematical function of position, or as a data cube. It is therefore possible to calculate output spectra for complex matter distributions expected from hydrodynamical simulations of the circumnuclear environment (e.g., \citealt{wada-2012}). However, for fitting limited-quality X-ray data, these structures need to be simplified and parametrized. For the spectral templates presented here, we choose the same simple, approximately toroidal geometry employed in \citet{brightman+nandra-2011a}. They used a uniform-density sphere with two conical polar cutouts with the opening angle as a free parameter of the model.\footnote{Specifically, this geometry is assumed for the \xspec\ table model \texttt{torus1006.fits}.} This simplification should be thought of as a smoothed distribution of individual clouds comprising the torus, as illustrated in \autoref{fig:sketch}. The adopted geometry implicitly assumes that any individual cloud is much smaller than the torus itself, and that clouds take up most of the approximately toroidal volume. Although we used the same geometry as \citet{brightman+nandra-2011a}, our new calculation is more detailed, flexible, includes features that the original calculation lacks, and resolves some known problems and errors (see, \citealt{liu+li-2014}). We directly compare our new spectral templates with those from \bntorus\ in \S\,\ref{sec:model-comparison}.

\borus\ calculates Green's functions for initial photon energies between 1\,keV and 1\,MeV. These functions are convolved with a parametrized intrinsic continuum in post-processing. The medium is assumed to be cold, neutral, and static. Photons are propagated through this medium until they are absorbed without fluorescent re-emission or until they escape the system. At each step, relative probabilities of photoelectric absorption and Compton scattering are computed based on photoelectric absorption cross-sections from the NIST/XCOM database\footnote{Available on the Web at \url{https://www.nist.gov/pml/xcom-photon-cross-sections-database}; originally \citet{xcom-original}.}, elemental abundances from \citet{anders+grevesse-1989}, and the Klein-Nishina scattering cross-section formula. In the case of absorption, fluorescent photons are emitted according to fluorescent yields for K$\alpha_1$, K$\alpha_2$ and K$\beta$ lines from \citet{krause-1979}, for all elements up to zinc ($A<31$).

Compton scattering for the spectral templates presented here does depend on atomic species, but we neglect the internal structure of the scattering atoms for the low-resolution templates discussed in this paper (see \citealt{furui+2016} for a calculation that includes these effects). The Compton shoulder is computed for all fluorescent spectral lines. Although our models are calculated on an energy grid with resolution sufficient for modeling of X-ray spectra from high-resolution calorimeter instruments, early versions of \xspec\ tables used in this paper have limited photon statistics and therefore lower energy resolution, sufficient for \nustar\ and CCD-based spectroscopy (see the left panel of \autoref{fig:repcomp_extra}). We compute spectral templates with a range of relative abundance of iron ($A_{\rm Fe}$) between 1/10 and 10. Changing the iron abundance parameter results in a self-consistent modification of iron fluorescent line intensity and of the shape of the reprocessed continuum, which is affected by the change in the total photoelectric cross-section. An example is given in the right panel of \autoref{fig:repcomp_extra}.

\vspace{2cm}

\subsection{\xspec\ Table Model \borustwo}
\label{sec:model-borus02}

The grid of spectral templates computed using \borus\ in the particular geometry shown in \autoref{fig:sketch} is named \borustwo. The covering factor of the torus, as seen from the X-ray source in the center, is simply related to the half-opening angle of the torus, \thetator, as \cftor\,$=\cos$\,\thetator. We note that this equality holds as long as clouds take up most of the torus volume; in cases where the space between dense clouds dominates the volume of the putative torus, the relation is more complicated (see, e.g., \citealt{nenkova+2008a}). Angles \thetator\ and \thetainc\ (inclination) both increase away from the axis of symmetry of the torus. We calculate the spectral templates for covering factors at 10 points equally spaced in $\cos$\,\thetator. The minimal and maximal values of \thetator, corresponding to the covering factors of 100\,\% and 10\,\%, are zero and $84.1^{\circ}$, respectively. The output of the radiative transfer simulation is arranged so that exit angles of each photon are separated into 10 bins in $\cos$\,\thetainc, each with a width of 0.1. The centers of the first and the last bins are at $\cos$\,\thetainc\ equal to 0.05 and 0.95, which corresponds to inclination angles of $87.1^{\circ}$ and $18.2^{\circ}$, respectively. Note that \thetainc\,$\approx 0^{\circ}$ corresponds to a pole-on and \thetainc\,$\approx 90^{\circ}$ corresponds to an edge-on view. Azimuthal angles are averaged over because of axial symmetry.

We utilize the additive table model option available in \xspec\ to enable fitting our parametrized grid of spectral templates to X-ray data. The FITS-format tables containing the spectra for the full range of parameters are named \texttt{borus02\_vYYMMDDx.fits}, where \texttt{YYMMDD} stand for the release date, and \texttt{x} marks a particular table version. For simplicity and convenience, we also make available versions of tables with a reduced number of parameters, with parameters in different units (\thetator, \thetainc, or their cosines), and with only line or only continuum emission.\footnote{In particular, the \texttt{borus01\_vYYMMDDx.fits} models represent a spherical absorber (covering factor fixed at unity, which is included in \borustwo\ tables). It can be directly compared to the \citet{brightman+nandra-2011a} model with the uniform sphere geometry (\xspec\ table \texttt{sphere0708.fits}), and the \texttt{plcabs} model \citep{yaqoob-1997}, which is a limited analytic approximation of radiative transfer in the same geometry. Note that the table naming scheme corresponds to a wider set of torus geometries computed using \borus, but not discussed in this paper.} The photon statistics of tables dated \texttt{170323}, which we use for fitting examples in \S\,\ref{sec:fitting}, are sufficient for analysis of \nustar\ data and medium-quality CCD-based soft X-ray spectra (FWHM\,$\gtrsim$\,130\,eV), as shown in \autoref{fig:repcomp_extra}. Their use on the highest-quality soft X-ray data or X-ray grating spectra (FWHM\,$<$\,130\,eV) is not recommended; future versions, however, will be adequate for such analyses. All \borustwo\ tables contain {\em only} the spectral components arising from reprocessing in the torus. The angular function of the transmitted line-of-sight component would be just a step function in the geometry assumed for this model. Such a component can be represented adequately by line-of-sight extinction models already available in \xspec. 

In the set of spectral templates presented in this paper, the intrinsic continuum is assumed to be a power law with an exponential cutoff, $n(E)\propto E^{-\Gamma}\exp (-E/E_{\mbox{\small cut}})$. The photon index ($\Gamma$) can be varied between 1.4 and 2.6, and the high-energy cutoff (\ecut) has a range between 20\,keV and 2\,MeV. Normalization of the intrinsic continuum follows the \xspec\ convention and is therefore defined in units of photons\,s$^{-1}$\,cm$^{-2}$\,keV$^{-1}$ at 1\,keV. These parameters can be linked to other basic spectral components in \xspec\ in order to construct a complete spectral model for fitting AGN X-ray spectra.

A basic model may be defined with the following command sequence in \xspec:
\begin{eqnarray*}
\texttt{m} &=& \texttt{c}_1\,*\,\texttt{phabs}\,*\,\left(\texttt{atable\{borus02\_vYYMMDx.fits\}}\right.\\
            &+& \texttt{zphabs}\,*\,\texttt{cabs}\,*\,\texttt{cutoffpl}\\
            &+& \left.\texttt{c}_2\,*\,\texttt{cutoffpl}\,\right)
\end{eqnarray*}
In the expression above, \texttt{c}$_1$ and \texttt{c}$_2$ stand for instrument cross-normalization and the relative normalization of a leaked or scattered unabsorbed reflection of the intrinsic continuum, respectively. \texttt{phabs} accounts for foreground Galactic absorption, while \texttt{zphabs}$\times$\texttt{cabs} represents line-of-sight absorption at the redshift of the X-ray source (generally independent from the average column density of the torus), including Compton scattering losses out of the line of sight.\footnote{The line-of sight absorption model \texttt{phabs} may be freely replaced with a more updated absorption model, such as \texttt{tbabs} \citep{wilms+2000}. Here we use \texttt{phabs} in order to consistently use elemental abundances adopted for calculation of the reprocessed spectra.} The \nh\ parameter of both \texttt{zphabs} and \texttt{cabs} needs to be the same in order to correctly account for total extinction along the line of sight. \texttt{cutoffpl} represents the intrinsic continuum, and its parameters should be linked to the $\Gamma$, \ecut, and normalization parameters of the \borustwo\ table.

We recommend formulating the model so that the \borustwo\ table is the first additive component. In that case the allowed parameter range for $\Gamma$ and \ecut\ will be read from the table, ensuring that in parameter optimization \xspec\ will not step out of the limited parameter space. The model can also be used in a setup with line-only and continuum-only tables, e.g., when one wishes to measure the flux of those components separately. In that case, the \texttt{atable} term in the definition of an \xspec\ model given above should be separated into a sum of the line and continuum components, with all their parameters linked.

The line-of-sight column density, \nhlos, and the torus column density, \nhtor, should generally not be linked -- the main feature of our new table model is that the equality of these two quantities can be tested with the data. However, the user may still choose to make the assumption that \nhtor\,$=$\,\nhlos\ in order to reduce the number of free parameters. For increased linearity of the parameter space, it is often better to use logarithmic units for \nhtor, fitting for \lognhtor\ instead. Likewise, we recommend fitting for the torus covering factor, \cftor\,$=\cos$\,\thetator, and the cosine of the inclination angle, instead of fitting for \thetator\ and \thetainc\ directly. Due to the likely complex landscape of the parameter space, the use of Markov chain Monte Carlo (MCMC) sampling or more advanced Bayesian methods \citep{buchner+2014} should be preferred over the straightforward $\chi^2$ minimization with many free parameters. In order to facilitate the application of our new \xspec\ table model to X-ray data, in \S\,\ref{sec:fitting} we present a two-step approach demonstrated on four AGN with \nustar\ data of different quality.

\begin{figure*}
\begin{center}
\includegraphics[width=0.95\textwidth]{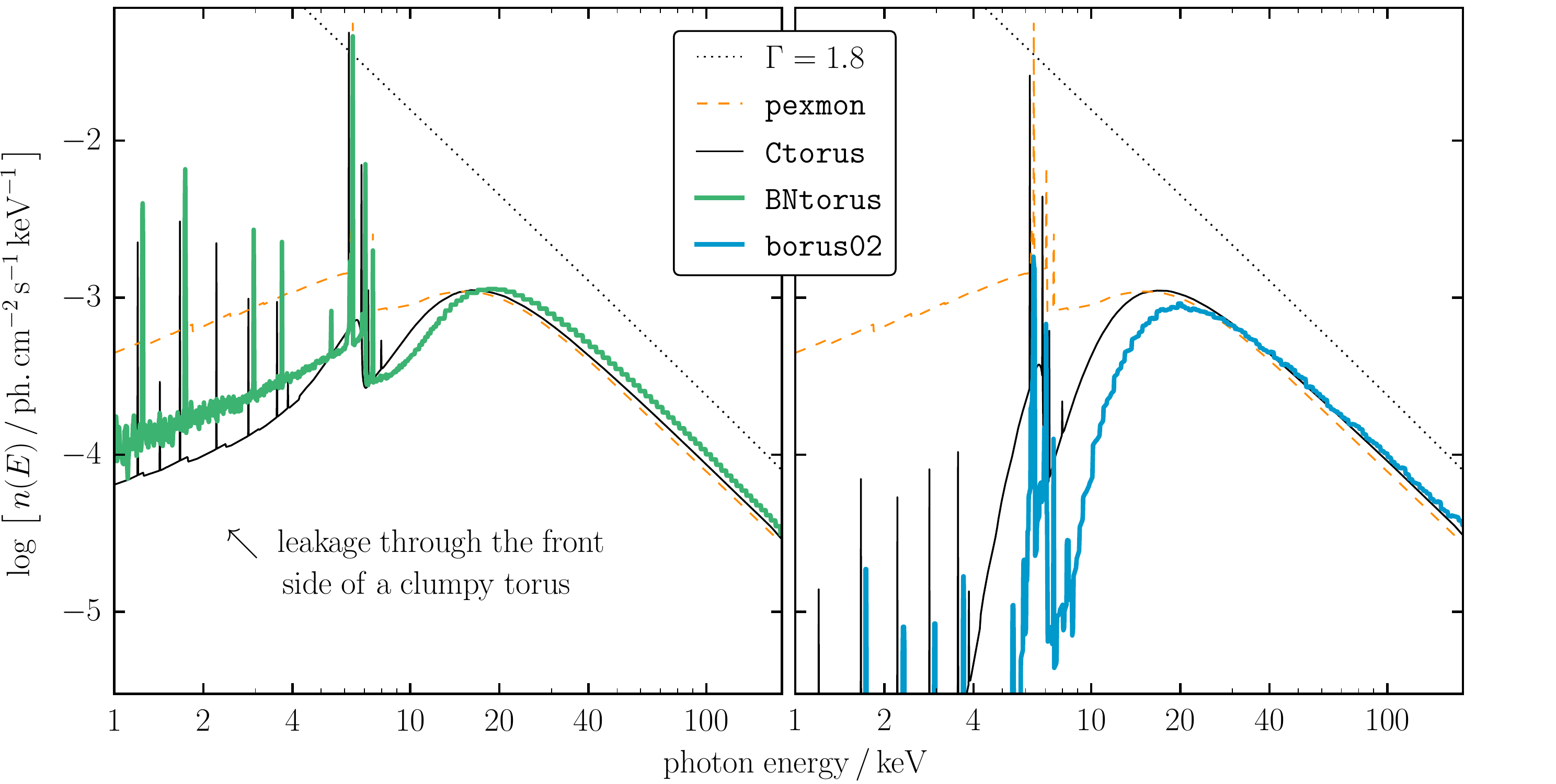}
\caption{Comparison of reprocessed spectra from \pexmon\ (\citealt{nandra+2007}; dashed orange line), \ctorus\ (\citealt{liu+li-2014}; thin, black, solid line), \bntorus\ (\citealt{brightman+nandra-2011a}; thick, green, solid line in the left panel), and \borustwo\ (thick, solid, blue line in the right panel). Intrinsic power-law continuum with $\Gamma=1.8$, assumed for each of the models, is shown with a grey, dashed line. For all models we assume an edge-on view (\thetainc\,$\approx 85^{\circ}$), and assume that the line-of-sight column density is equal to the average column density of the torus (\lognhloscmmt\,$=$\,\lognhtorcmmt\,$=24.5$). For comparison, \pexmon\ is plotted with normalization formally corresponding to a covering factor of 50\,\% (\rpex\,$=1$; infinite plane geometry), and all other models have a 50\,\% covering factor (\thetator\,$=60^{\circ}$). In the left panel, \ctorus\ is plotted with a small number of clouds in the line of sight along the equatorial plane ($N_{\mbox{\small clo}}=2$). \bntorus, also plotted in the left panel, matches this porous torus model relatively well. In the right panel, we compare \borustwo\ to \ctorus\ with a larger number of clouds ($N_{\mbox{\small clo}}=10$), approximating a more uniform torus. In reference to \autoref{fig:sketch}, the right and left panel mainly show the effects of absorption and scattering between points D and E, and lack thereof, respectively. \label{fig:modcomp} }
\end{center}
\end{figure*}

\subsection{Direct Comparison with \bntorus}
\label{sec:model-comparison}

In this section we highlight the differences between our new set of X-ray spectral templates and the frequently used, publicly available \xspec\ table model \bntorus. The physics of radiative transfer employed in both calculations is nearly the same; the \borus\ code is more versatile in terms of geometry, operates at higher energy resolution, and takes into account a greater number of atomic species and their fluorescent lines. For \borustwo\ we adopted the same approximately toroidal geometry assumed in calculating the \bntorus\ spectral templates, at least in principle.

As \citet{liu+li-2015} recently pointed out, there is significant disagreement between \bntorus\ and their simulations for the same geometry. Their \ctorus\ model qualitatively agrees with \mytorus\ and \etorus\ models, although direct comparisons are not straightforward due to different assumed geometries (see the Appendix for more details). These models, as well as previous calculations by \citet{ghisellini+1994} and \citet{krolik+1994}, suggest that \bntorus\ produces a significant excess of soft X-ray flux at nearly edge-on inclination. A comparison of \bntorus\ spectra to our new calculations (see \autoref{fig:modcomp}) confirms such a discrepancy. We trace the problem back to an error in the original calculation of the \bntorus\ model.

The issue with the \citet{brightman+nandra-2011a} calculation arises from absorption not being applied to the reprocessed light emitted from the inner side of the torus (the side opposite the observer). All obscured sightlines are affected by this to some degree. The photon path shown with white dashed lines in \autoref{fig:sketch} exemplifies the issue: by error, photons scattered toward an edge-on observer at point C (near the inner surface of the torus) reach the observer without any further absorption or scattering. In the assumed geometry, these photons should additionally interact with the torus material between points D and E. As a result of the missing absorption, within the \bntorus\ model there is very little difference in the spectral shapes of reprocessed components for pole-on and edge-on inclinations.

The disagreement between \bntorus\ and our new calculation is demonstrated in \autoref{fig:modcomp}. In order for the reprocessed component to dominate below $\simeq$30\,keV, we compare model spectra for \lognhloscmmt\,$=$\,\lognhtorcmmt\,$=24.5$. The \bntorus\ spectrum shows an excess of soft X-ray flux ($\lesssim$20\,keV), which should be heavily absorbed for an edge-on view of a uniform-density torus. We further compare the spectra to the clumpy \ctorus\ model, which features the average number of clouds along and equatorial line of sight ($N_{\mbox{\small clo}}$, ranging from 2 to 10) as a free parameter. For $N_{\mbox{\small clo}}=2$, it emulates a torus sparsely populated with clouds, which results in less absorption and scattering on the side of the torus closer to the observer. In the other extreme, for $N_{\mbox{\small clo}}=10$, the torus volume is filled out more and therefore more similar to a uniform-density torus. The former situation matches \bntorus\ well, while the latter is closer to \borustwo. \autoref{fig:modcomp} shows the difference only for \lognhtorcmmt\,$=24.5$; for higher and lower \nhtor, the differences are more and less severe, respectively.

While \bntorus\ does not correctly reproduce spectra for the geometry described in \citet{brightman+nandra-2011a}, it may approximate spectral features produced by the more general and more realistic class of clumpy tori. In the example given in \autoref{fig:sketch}, the photon path passes the cross-section of the torus (between points D and E) through a region with no clouds. This is a physically plausible scenario in which photons emitted from the inner side of the torus would be able to escape unimpeded toward an edge-on observer. The inner-side reprocessed component, which shows some similarity to that reproduced by \bntorus\ (see \autoref{fig:modcomp}), could in principle be directly observable through the front side of the torus as long as the gas distribution is not uniform, i.e., is clumpy. Evidence that this is a possible, if not likely, scenario in AGN is abundant from detailed spectroscopy (e.g., \citealt{arevalo+2014-circinus}, \citealt{balokovic+2014}, \citealt{annuar+2015-ngc5643}) and studies of line-of-sight absorption variability (e.g., \citealt{risaliti+2002}, \citealt{torricelliCiamponi+2014}, \citealt{markowitz+2014}). A more detailed, quantitative assessment of the error introduced by \bntorus, which could lead to deducing a correction factor for existing results, is a complex task, and will be the aim of future work.

In addition to resolving the issue of missing absorption, \borustwo\ supersedes \bntorus\ with additional features that make it significantly more flexible. Separation of the line-of-sight and reprocessed components is not possible with \bntorus. This limits its application only to AGN for which it is justified to assume that the line-of-sight column density (\nhlos) is equal to the average column density of the torus (\nhtor). With \borustwo\ one can self-consistently model multi-epoch data assuming that the \nhlos\ varies, while \nhtor\ does not, as observed in many AGN with multi-epoch X-ray data (e.g., \citealt{marchese+2012-ngc454}, \citealt{braito+2013-ngc4507}, \citealt{ricci+2016-ic751}). This important feature allows us to test commonly made assumptions regarding \nhtor\ and \nhlos, and to more directly probe the structure of AGN obscuration. Furthermore, \borustwo\ includes the high-energy cutoff (\ecut) and the relative abundance of iron ($A_{\mbox{\small Fe}}$) as additional model parameters. They enable complex spectral models to be used with a greater degree of self-consistency. These parameters will also make it possible to include torus reprocessing components in spectral models of AGN in which these parameters appear to have extreme values (e.g., \citealt{brenneman+2011-ngc3783}, \citealt{kara+2015-1h0707}, \citealt{xu+2017-iras05189}).

\begin{figure}[t!]
\begin{center}
\includegraphics[width=0.98\columnwidth]{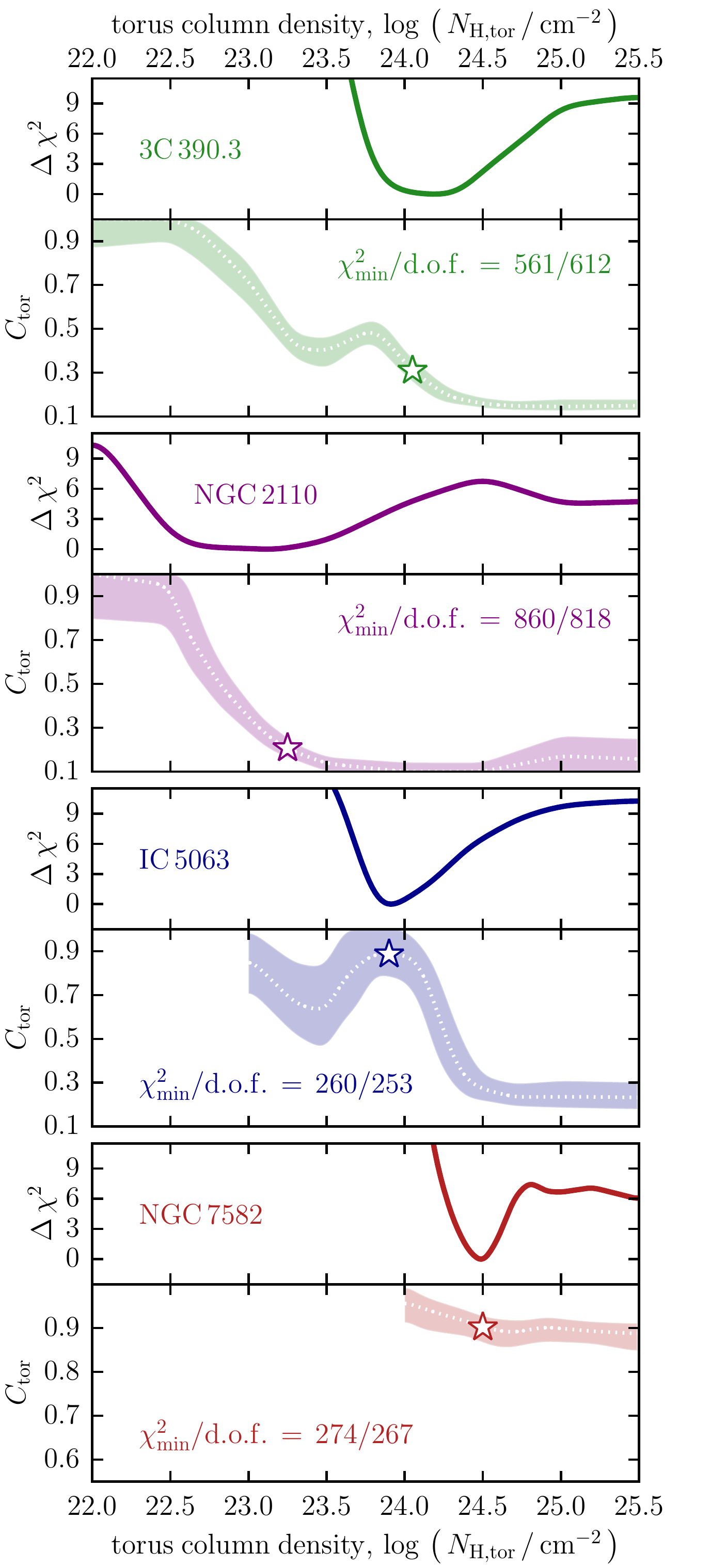}
\label{fig:sweep}
\caption{Constraints on the torus covering factor (\cftor) as a function of the torus column density (\nhtor) for 3C\,390.3, NGC\,2110, IC\,5063 and NGC\,7582. Inclination is a free parameter in all fits. The solid lines in the upper panels for each source show minimum $\chi^2$ as a function of \nhtor\ normalized to the best-fit $\chi^2$. The dashed lines in the lower panels show medians of the probability distribution of \cftor\ at each \nhtor\ and the shaded regions enclose 68\,\% (1\,$\sigma$) of the probability. Curves are plotted only for the range of \nhtor\ for which a statistically acceptable fit was found. They have been smoothed with a Gaussian kernel with standard deviation equal to half of the step size over the parameter space, $\Delta$\,\lognhtorcmmt\,$=0.1$. Stars mark the \nhtor\ with the lowest $\chi^2$ for each AGN.}
\end{center}
\end{figure}

\section{Examples of Application for\\Fitting Hard X-ray Spectra}
\label{sec:fitting}

In order to demonstrate the usage and potential of our model, we choose four AGN observed with \nustar: 3C\,390.3, NGC\,2110, IC\,5063 and NGC\,7582. This is by no means a complete or representative sample -- the targets are primarily chosen for the diversity of their physical properties. 3C\,390.3 is a broad-line radio galaxy (a radio-loud type\,1 Seyfert), IC\,5063 is a radio-loud type~2 Seyfert, and NGC\,2110 and NGC\,7582 are radio-quiet type~2 Seyferts. Except for IC\,5063, which is part of a large sample presented in B18 (as are NGC\,2110 and NGC\,7582), detailed spectral analyses of the \nustar\ spectra of these sources have already been published: 3C\,390.3 by \citet{lohfink+2015-3c390d3}, NGC\,2110 by \citet{marinucci+2015-ngc2110} and NGC\,7582 by \citet{rivers+2015-ngc7582}. In the case of NGC\,7582, we additionally include a new \nustar\ observation taken in 2016 (obsID 60201003002), which has not yet been published elsewhere. The reader is referred to the references listed above for the description of the observations, data processing procedures, and spectral analyses using spectral models commonly employed in the literature. For simplicity, in this paper we choose to use only the \nustar\ data (3--79\,keV) for fitting our model.

We performed spectral analyses in \xspec, fitting FPMA and FPMB spectra simultaneously, without coadding. Our basic model is defined as in \S\,\ref{sec:model-borus02}, but with the factor {\texttt{c}$_2$ (relative normalization of the secondary continuum) fixed to zero, and \ecut\ fixed to 300\,keV, unless explicitly stated otherwise. Parameter optimization is based on the $\chi^2$ statistic. We use 5\,\% as the threshold in null-hypothesis probability (\pnull; the probability of the observed data being drawn from a particular model, given its $\chi^2$ and the number of degrees of freedom) for a model to be formally acceptable as a good representation of the data; i.e., models with \pnull\,$<5$\,\% are rejected. We quote uncertainties on the fitted model parameters based on marginalized probability distributions derived from converged MCMC chains produced with the built-in \citet{goodman+weare-2010} MCMC algorithm in \xspec. Uncertainty is quoted as the interval containing 68\,\% of the total probability, equivalent to 1\,$\sigma$ uncertainty. When this interval includes the edge of the finite parameter domain, we quote a 1\,$\sigma$ constraint, so that 84\,\% of the total probability is enclosed (conversely, 16\,\% is left out). We verified that the best-fit parameters are always within the uncertainty interval, although they often do not exactly match the distribution medians.

In \S\,\ref{sec:fitting-single} we first present results based on a single epoch of \nustar\ data for each source (the first epochs for NGC\,2110 and NGC\,7582). However, single-epoch spectral fits may be biased by the temporary increase or decrease of the intrinsic continuum that is not accompanied by a corresponding change in the reprocessed component due to the extended nature of the torus. In \S\,\ref{sec:fitting-extra} we show how single-epoch constraints may be influenced by variability, and discuss how multi-epoch X-ray data and some justifiable assumptions can be leveraged to derive more robust self-consistent constraints and assess possible systematics.

\begin{figure*}[t!]
\begin{center}
\label{fig:eeuf}
\includegraphics[width=1.0\textwidth]{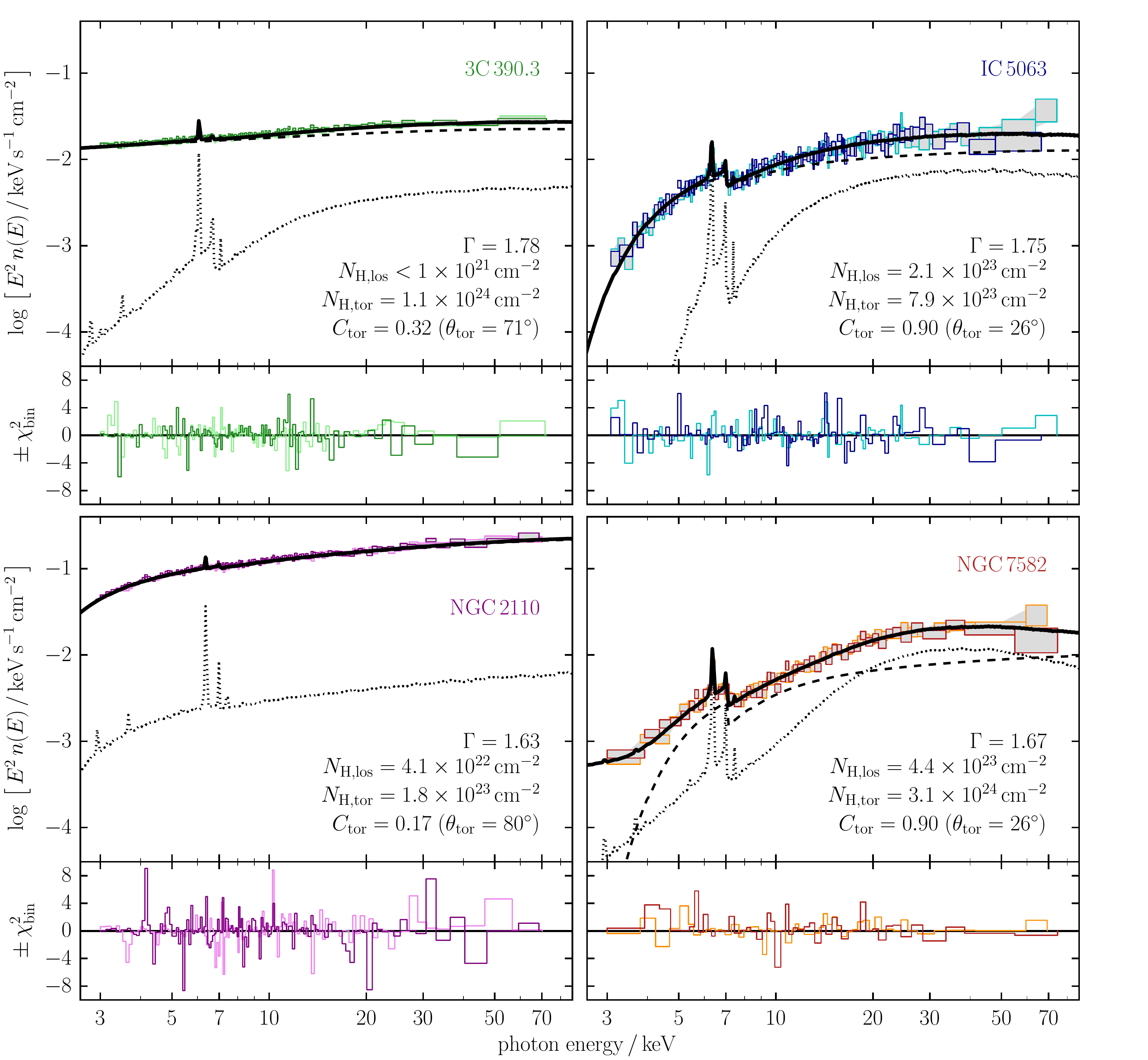}
\caption{Best-fit spectral models for 3C\,390.3, NGC\,2110, IC\,5063 and NGC\,7582. We show the total model spectrum in solid lines, the absorbed power-law components with dashed lines, and the reprocessed component (including fluorescent emission lines) with dotted lines. In the lower right of each panel showing the model spectra, we list its main parameters. FPMA and FPMB residuals in terms of $\chi^2$ contributions per bin are plotted below each spectrum (in darker and lighter colors, respectively); they are binned to improve clarity.}
\end{center}
\end{figure*}

\subsection{Single-epoch Constraints on the Torus Parameters}
\label{sec:fitting-single}

We first run a set of fits with the \nhtor\ parameter kept fixed in order to determine whether any assumptions on this parameter result in statistically unacceptable solutions (i.e., \pnull\,$<5$\,\%). The pairs of panels in \autoref{fig:sweep} show the constraints on \cftor\ for a range of assumed values of \nhtor, and the associated $\chi^2$ curve. Other model parameters are left free to vary in these fits. Spectral models for each of the sources are shown in \autoref{fig:eeuf}. In \autoref{fig:nh_cf_plane} we show two-dimensional constraints on \nhtor\ and \cftor, where \nhtor\ is also left free to vary in the fit.

\subsubsection{3C\,390.3}
\label{sec:fitting-single-3c390p3}

The unobscured 3C\,390.3 is our simplest example, since the lack of line-of-sight absorption in the \nustar\ band allows us to set \nhlos\ to zero. The top two panels in \autoref{fig:sweep} show that a good fit can be found for any assumed \nhtor, but that there is a clear minimum in $\chi^2$ around the best fit at \nhtor\,$=1.1\times 10^{24}$\,\cmmt. For the best fit, $\chi^2=561.0$ for 612 degrees of freedom (d.o.f.). This model is shown in the top left panel of \autoref{fig:eeuf}; the flat residuals suggest that all features in the data are accounted for by this spectral solution. With \nhtor\ fixed at the best-fit value, we find \cftor\,$=0.32_{-0.07}^{+0.05}$. The inclination angle is constrained to $\cos$\,\thetainc$>0.3$, so that our line of sight does not intercept any of the reprocessing material, consistent with the type 1 optical classification of 3C\,390.3. \autoref{fig:nh_cf_plane} shows the probability density distribution in the two-dimensional plane spanned by \nhtor\ and \cftor, obtained from a fit in which \nhtor\ is left free to vary. In this case, the covering factor constraint is slightly broader: \cftor\,$=0.3_{-0.1}^{+0.2}$. The possibility that the reprocessed component is due to the accretion disk rather than the torus is discussed in \S\,\ref{sec:fitting-extra-comp}.

\subsubsection{NGC\,2110}
\label{sec:fitting-single-ngc2110}

NGC\,2110 is mildly obscured by \nhlos\,$=(4.1\pm0.2)\times 10^{22}$\,\cmmt, which is detectable as an exponential roll-off of the power-law continuum ($\Gamma=1.628\pm0.007$) at the lower end of the \nustar\ band. Its spectrum is remarkably featureless, except for a narrow Fe\,K$\alpha$ line with an equivalent width of $33\pm6$\,eV. Stepping through the range of \lognhtorcmmt\ between 22.0 and 25.5 we find that acceptable model include tori with very small covering factors, \cftor\,$<0.2$ for \lognhtorcmmt\,$>23.5$, as well as tori with high covering but low \nhtor. The difference in the best-fit $\chi^2$ over the whole range is very small ($<\!10$, for 819 d.o.f.). A broad minimum in $\chi^2$ at $22.5<$\,\lognhtorcmmt\,$<23.5$ covers nearly the full range of covering factors (0.1--1.0). The best fit, with $\chi^2/$d.o.f.$=860.4/818$, is found for \lognhtorcmmt\,$\approx 23.3$. With \lognhtor\ fixed at this value, \cftor\,$<0.24$ and $\cos$\,\thetainc\,$>0.28$. The lower left pair of panels in \autoref{fig:eeuf} show this model and the residuals. Note that a number of narrow, isolated bins contribute substantially to the total $\chi^2$ without corresponding to any real but unmodeled spectral features. With \nhtor\ as a free parameter in the fit, the constraints are much broader, as shown in \autoref{fig:nh_cf_plane}. The probability density distribution in the \nhtor--\cftor\ plane is highly elongated and reaches \lognhtorcmmt\,$\approx 22.6$ and \cftor\,$\approx 0.8$ within 1\,$\sigma$ contours. No constraint on inclination can be given in this case.

\subsubsection{IC\,5063}
\label{sec:fitting-single-ic5063}

Partly due to higher line-of-sight absorption in comparison with 3C\,390.3 and NGC\,2110, the \nustar\ data for IC\,5063 have the constraining power to reject a part of the parameter space on statistical grounds, despite lower photon statistics. As the third pair of panels in \autoref{fig:sweep} shows with the lack of \cftor\ constraints for \nhtor\,$<10^{23}$\,\cmmt, no model with \pnull\,$>5$\,\% can be found for a lower torus column density. $\chi^2$ as a function of \lognhtor\ has a minimum ($\chi^2/$d.o.f.$=259.6/253$) around \lognhtorcmmt\,$=23.9$. With \lognhtor\ fixed at this value, \cftor\,$>0.77$ and $\cos$\,\thetainc\,$<0.62$. This model is shown in the upper right panels of \autoref{fig:eeuf}. 
Fitting for the torus column density, we find that it is very well constrained, \lognhtorcmmt\,$=23.95\pm0.07$, and that there is almost no degeneracy with the covering factor. Unlike the cases of 3C\,390.3 and NGC\,2110, the contours in the \nhtor--\cftor\ plane are elongated along the axes, predominantly vertically. Constraints on \cftor\ and $\cos$\,\thetainc\ are therefore no different than those obtained with \nhtor\ fixed. With a high \cftor\ and \nhtor\ near the CT threshold, the reprocessed component contributes $\sim20$\,\% of the flux in the 10--50\,keV band.

\subsubsection{NGC\,7582}
\label{sec:fitting-single-ngc7582}

NGC\,7582 exhibits the most complex X-ray spectrum of the AGN discussed here. Its \nhlos\ is known to be variable and multiple layers of absorption have been invoked in previous spectral analyses \citep{rivers+2015-ngc7582}. We find it necessary to include a non-zero parameter \texttt{c}$_2$ (as defined in \S\,\ref{sec:model-borus02}) in order to account for partial absorption along the line of sight; without it, the residuals show a significant excess below 4.5\,keV. For the first \nustar\ observation considered here, under different assumptions for \nhtor, we always find \nhlos\ consistent with $(3.6\pm0.4)\times 10^{23}$\,\cmmt\ and \texttt{c}$_2=0.10\pm0.04$ (i.e. $\approx90$\,\% line-of-sight covering, or $\approx10$\,\% Thompson-scattered fraction). The 3--15\,keV continuum is dominated by the transmitted component, while the Compton hump dominates in the 15--60\,keV range (see the lower right panels of \autoref{fig:eeuf}). Stepping through the \nhtor\ parameter space we first find that no models with \lognhtorcmmt\,$<24.0$ are acceptable according to our \pnull\,$>5$\,\% threshold. The $\chi^2$ curve shown in the lowest pair of panels in \autoref{fig:sweep} shows a very well defined minimum at \lognhtorcmmt\,$\approx24.5$. At this \nhtor, both \cftor\ and $\cos$\,\thetainc\ are narrowly constrained to $0.90\pm0.03$. Additionally fitting for the torus column density yields \lognhtorcmmt\,$=24.0\pm0.1$, and does not affect the other model parameters. NGC\,7582 therefore seems to have a CT torus that covers $\approx90$\,\% of the sky as seen from the SMBH, yet we observe it through a hole with roughly an order of magnitude lower column density. We further test this result with additional data in the following section, and discuss its interpretation in \S\,\ref{sec:discussion}.

\subsection{Additional Constraints and Considerations}
\label{sec:fitting-extra}

\subsubsection{Line-of-sight and Torus Column Density}
\label{sec:fitting-extra-nhtest}

The flexibility of \borustwo\ allows us to test the common assumption that the line-of-sight column density matches the average column density of the torus. For 3C\,390.3 and NGC\,7582, with \lognhloscmmt\,$<21$ and \lognhloscmmt\,$\approx23.3$, respectively, this assumption cannot yield a good fit for any combination of other model parameters. Both AGN clearly require presence of CT material out of our line of sight. This is not necessarily true for IC\,5063 and NGC\,2110, since statistically acceptable models with \pnull\,$>5$\,\% can be found for both AGN. For NGC\,2110, such a solution (\lognhtorcmmt\,$=$\,\lognhloscmmt\,$\approx22.6$, \cftor\,$>0.7$) is within the 1\,$\sigma$ contour for the single epoch constraints shown in \autoref{fig:nh_cf_plane}, and within the 3\,$\sigma$ contour based on two epochs. Fitting the IC\,5063 data with the assumption that \nhtor\,$=$\,\lognhloscmmt\,$\approx23.3$ increases $\chi^2$ with respect to the fit featuring independent column densities ($\Delta \chi^2 = 16.0$) and results in \cftor\,$=0.5\pm0.1$. In this case, we also find $\Gamma=1.51\pm0.03$, which implies a harder intrinsic continuum than the bulk of local Seyferts (for which the distribution of $\Gamma$ is roughly Gaussian with a mean $\simeq$1.8 and standard deviation $\simeq$0.2; e.g., \citealt{dadina-2008}, \citealt{rivers+2013}, B18), unlike $\Gamma=1.75\pm0.04$ obtained in \S\,\ref{sec:fitting-single}. 

\vspace{0.1cm}
\subsubsection{Additional Spectral Parameters and Components,\\ and External Constraints}
\label{sec:fitting-extra-comp}

Model parameters \ecut\ (the high-energy cut-off in the intrinsic continuum) and \afe\ the (relative abundance of iron) have been kept constant in the analysis thus far. However, letting these parameters vary in the fitting does not lead to significantly better fits, while it does result in additional degeneracy, i.e., in poorer constraints on other model parameters. The data considered in this paper do not show preference away from the assumed values \ecut$=300$\,keV and \afe$=$\afesun. The largest deviations we find are \afe/\afesun$=0.92\pm0.03$ for NGC\,7582, which does not shift other parameters by more than their 1\,$\sigma$ uncertainties, and \ecut$=155_{-8}^{+11}$\,keV for 3C\,390.3, which indicates a minor shift in the torus parameters (\lognhtorcmmt\,$=24.4\pm0.3$ and \cftor\,$=0.2\pm0.1$). However, the decrease in $\chi^2$ with respect to the fixed values is too small to consider these features significant ($\Delta\chi^2\lesssim3$). Note that this was not the case in the analysis of \citet{lohfink+2015-3c390d3}, where the reprocessed continuum was assumed to have a different spectral shape. 

Because 3C\,390.3 is a powerful radio galaxy, the orientation of its jet can be measured in order to better constrain the inclination (e.g., \citealt{alef+1988}). Assumption of co-alignment can then be employed to infer the inclination of the accretion disk and the torus. Based on this and constraints from other measurements, which yield similar values (e.g., \citealt{flohic+eracleous-2008}), \citet{dietrich+2012} found that the inclination of the symmetry axis of the AGN to our line of sight is $27\pm2$ degrees. If we fix the inclination and perform the fitting with \ecut\ free to vary, we obtain \lognhtorcmmt\,$>24.5$ and \cftor\,$=0.14\pm0.02$. While these constraints still marginally overlap with those obtained with \thetainc\ as a free parameter, the two-dimensional probability distribution is shifted appreciably toward higher \nhtor\ and lower \cftor. In \autoref{fig:nh_cf_plane}, we show this as an example of how different assumptions may systematically shift constraints on the torus parameters.

As 3C\,390.3 is a type~1 AGN, we also tested its spectrum for the presence of relativistically broadened reprocessing in the innermost part of the accretion disk by adding a \texttt{relxill} component \citep{garcia+2014} to our \xspec\ model. Over a variety of assumptions for the parameters of the \texttt{relxill} component, which we kept constant and consistent with the analysis of \citet{lohfink+2015-3c390d3}, we find that its contribution to the iron line emission and the Compton hump is always sub-dominant. In all cases, \cftor\ is found to be consistent within 2\,$\sigma$ with the region outlined by contours in the two panels of \autoref{fig:nh_cf_plane}. This brief analysis indicates that the reprocessed component is dominated by material which may have a disk-like geometry, but is located at distances not affected by general relativistic effects. We agree with the analysis of \citet{lohfink+2015-3c390d3}, which made use of \suzaku\ data with higher energy resolution, and only found evidence for distant reprocessing without any relativistically broadened features.

\begin{figure*}
\begin{center}
\includegraphics[width=1.0\textwidth]{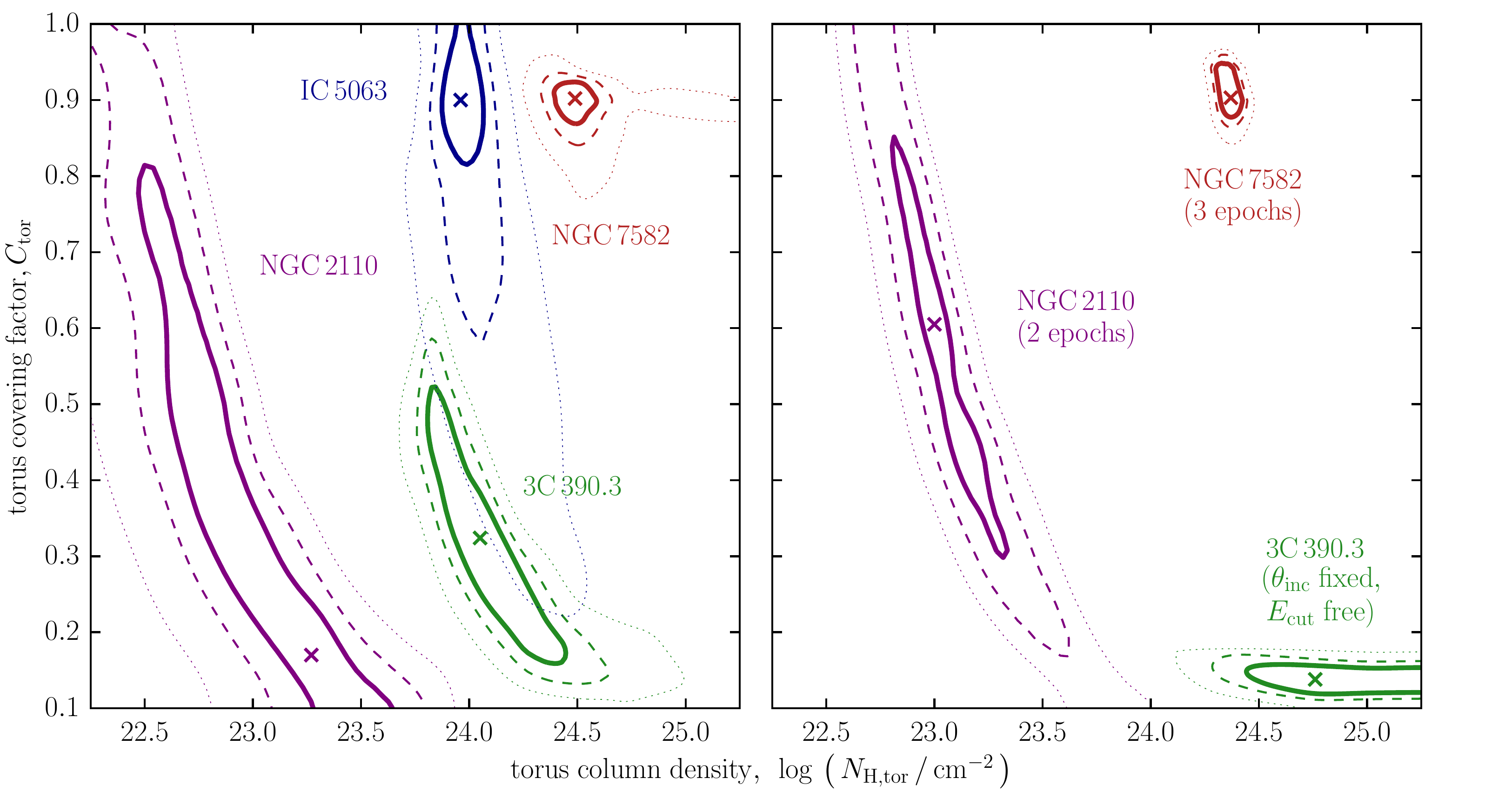}
\caption{Two-dimensional $\Delta \chi^2$ contours for torus column density and covering factor for 3C\,390.3, IC\,5063, NGC\,2110 and NGC\,7582. The left panel shows basic single-epoch fits with inclination as a free parameter and fixed \ecut\,$=300$\,keV. The right panel shows contours for the three AGN with additional constraints introduced in \S\,\ref{sec:fitting-extra}: two and three epochs fitted simultaneously for NGC\,2110 and NGC\,7582, respectively, and fixed inclination, \thetainc\,$=27^{\circ}$ \citep{dietrich+2012}, and fitted high-energy cutoff, \ecut\,$=115_{+11}^{-8}$\,keV, for 3C\,390.3. Solid, dashed and dotted lines mark 1, 2 and 3\,$\sigma$ contours ($\Delta\chi^2=2.3$, $4.6$ and $9.2$ from the best-fit value). Crosses mark the best-fit values. \label{fig:nh_cf_plane} }
\end{center}
\end{figure*}

\begin{figure*}
\begin{center}
\includegraphics[width=1.0\textwidth]{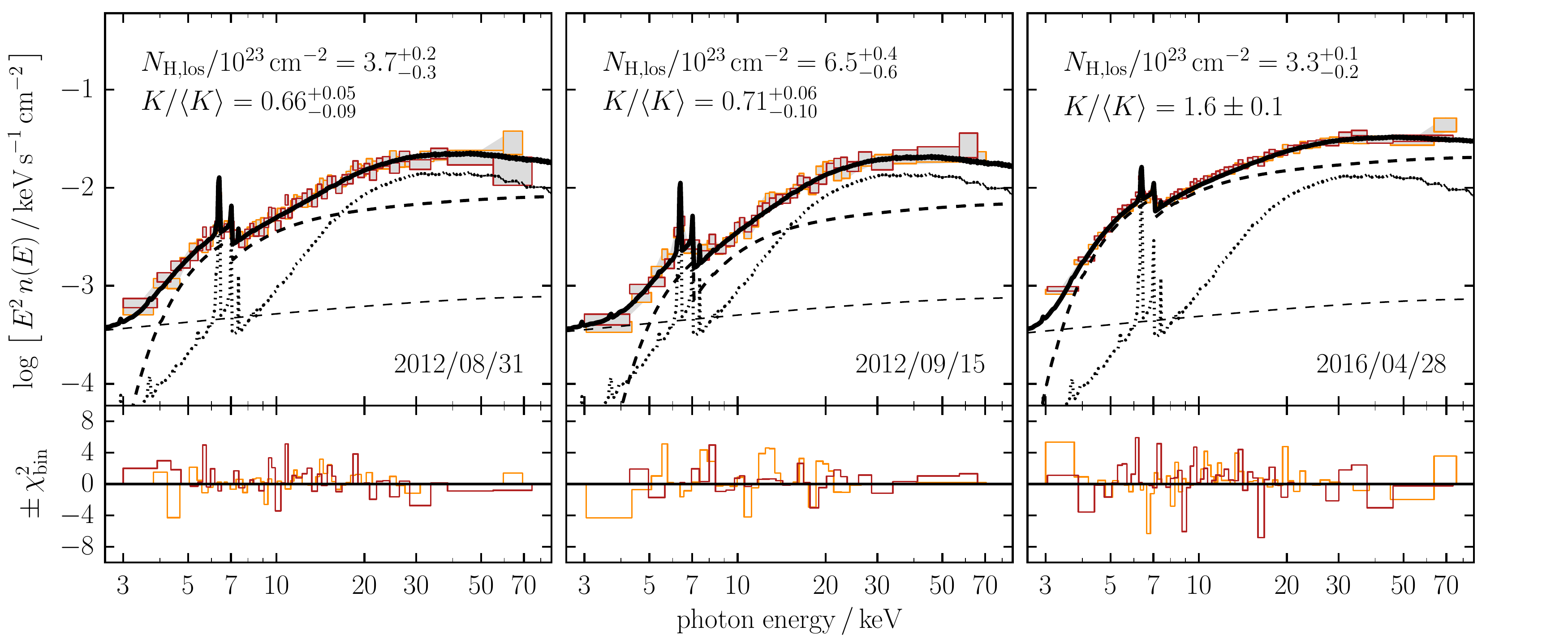}
\caption{Joint modeling of the three \nustar\ epochs of NGC\,7582 observation. The spectra can be self-consistently modeled under the assumption that the normalization of the intrinsic continuum ($K$) and the line-of-sight column density (\nhlos) vary between observations, and that the scattered components' normalization is determined by the average intrinsic continuum normalization ($\langle K \rangle$). See \S\,\ref{sec:fitting-extra-multi} for details. \label{fig:ngc7582_3epochs} }
\end{center}
\end{figure*}

\vspace{0.1cm}
\subsubsection{Multi-epoch X-ray Data}
\label{sec:fitting-extra-multi}

NGC\,2110 and NGC\,7582 have been observed with \nustar\ two and three times, respectively.\footnote{3C\,390.3 has formally been observed twice, but because those observations are consecutive, we treat them as a single observation here.} The advantage of multi-epoch observations is that the effects of variability in luminosity or other spectral components can be taken into account self-consistently. AGN are known to vary in luminosity of the intrinsic continuum down to very short timescales. However, the torus is expected to be a parsec-scale structure and hence the reprocessed spectral components cannot follow fast changes in the intrinsic continuum. The reprocessed components should therefore be normalized not with respect to the intrinsic continuum luminosity within a given observation, but with respect to the average luminosity. Multiple observations provide additional photon statistics that reduce statistical uncertainties, and they also provide a better estimate of the average, rather than instantaneous intrinsic luminosity of the AGN.

For NGC\,2110, a joint fit of two epochs yields results similar to those from our single-epoch analysis in \S\,\ref{sec:fitting-single}, with \lognhtor\,$=23.0\pm0.3$, \cftor\,$=0.6_{-0.3}^{+0.2}$, and no constraint on \thetainc. The two-dimensional probability distributions shown in \autoref{fig:nh_cf_plane} seem marginally inconsistent with each other. However, it must be noted that the two-epoch analysis is significantly more robust; not only does it have better photon statistics, but it also avoids the erroneous normalization of the reprocessed components with an intrinsic continuum that is atypically luminous.

In addition to the variability of the intrinsic continuum luminosity, some AGN, like NGC\,7582, also vary in the line-of-sight column density. In modeling multiple epochs of observation of such an object, we therefore allow for \nhlos\ to be fitted to each observation in addition to the intrinsic continuum amplitude. We assume that a good representation of the average intrinsic continuum luminosity, which sets the normalization of the reprocessed components, is provided by the average of continuum luminosities between the three observations. We found no evidence that the photon index changed between observations, while \nhlos\ and intrinsic continuum amplitude did.

We show our best-fit spectral model for NGC\,7582 in all three epochs of \nustar\ observations in \autoref{fig:ngc7582_3epochs}. Each of the three panels lists \nhlos\ and the continuum normalization with respect to the mean ($\langle K \rangle$). It is worth noting that in the first two epochs the Compton hump dominates the 15--60\,keV band, while in the third epoch the increased intrinsic continuum dominates instead. This self-consistent three-epoch fit confirms our torus constraints from \S\,\ref{sec:fitting-single}, and makes them even tighter: \lognhtorcmmt\,$=24.39\pm0.06$, \cftor\,$=0.90_{-0.03}^{+0.07}$, and $\cos$\,\thetainc\,$=0.87\pm0.05$. Interestingly, \citet{rivers+2015-ngc7582} have already found tentative evidence for a covering factor of 80--90\,\% and torus column density of $\sim3\times10^{24}$\,\cmmt\ using \mytorus. This was estimated from the normalization ratio of the intrinsic and the reprocessed continuum components when \mytorus\ was used in its decoupled configuration.

\vspace{0.5cm}
\section{Discussion}
\label{sec:discussion}

\subsection{New Tool for Studying the AGN Torus}
\label{sec:discussion-new}

In recent years, \mytorus, \bntorus, and, to a more limited extent, the torus model by \citet{ikeda+2009}, have been used to probe the basic parameters of AGN tori---their average column densities (\nhtor) and covering factors (\cftor)---from the hard X-ray band. The average column density can be estimated using the \mytorus\ model assuming its particular geometry with a fixed 50\,\% covering factor (e.g., \citealt{braito+2013-ngc4507}, \citealt{balokovic+2014}, \citealt{yaqoob+2015-mrk3}), and the covering factor can, in some cases, be estimated by giving up its self-consistency (e.g., \citealt{rivers+2015-ngc7582}). \bntorus\ has been used to provide covering factor estimates in previous studies (e.g., \citealt{gandhi+2014-mrk34}, \citealt{brightman+2015}, \citealt{koss+2015-ngc3393}); they may need reassessment in the light of issues with missing absorption that we discussed in \S\,\ref{sec:model-comparison}, which we leave for future work. Our new model is the only reliable publicly available tool for constraining the torus covering factor from X-ray data, as long as the approximation of a uniform density torus reasonably represents physical reality.

More complex torus models will become available in the near future (e.g., \citealt{liu+li-2014}, \citealt{furui+2016}, \citealt{paltani+ricci-2017}), motivated by the high energy resolution of X-ray calorimeters. The \borustwo\ table used in this paper (version \texttt{170323a}) is limited to low energy resolution by photon statistics. While this is sufficient for fitting \nustar\ data (see \autoref{fig:repcomp_extra} for an example), it is inadequate for analyses of the highest-quality CCD-based spectra or X-ray grating spectroscopy; however, future versions will feature better photon statistics and enable analyses with higher energy resolution.

As an updated and extended version of the already popular \bntorus\ model, \borustwo\ may be an effective tool for better understanding the relation between new results and those already in the literature. The \borus\ radiative transfer code, on which our \borustwo\ spectral templates are based, is a versatile tool for investigating the observable effects of torus geometry and clumpiness in the X-ray band in future studies (Balokovi\'{c} et al., in prep.). The parametrization of geometry adopted for \borustwo\ was chosen in particular to match the \bntorus\ model, in order to extend its flexibility and enable more detailed studies of torus parameters in a wider population of AGN than previously possible.

In terms of additional model parameters, \borustwo\ tables include the high-energy cutoff and the relative abundance of iron. More importantly, it combines the features of both \bntorus\ and \mytorus\ by having a variable covering factor, as well as the reprocessed component separated from the transmitted (absorbed) component. The line-of-sight column density therefore does not need to be assumed equal to the average over the whole torus; this common assumption can be tested with the data. This important feature enables self-consistent modeling of the torus for both unobscured and obscured AGN, including those showing variability in \nhlos. The clumpy torus scenario in which \nhlos\ and \nhtor\ generally differ is supported by the \nustar\ data in the fitting examples presented in \S\,\ref{sec:fitting}, as well as the literature.

\vspace{0.5cm}
\subsection{Interpretation of Fitting Results}
\label{sec:discussion-interp}

In interpretation of the results from spectral analyses employing \borustwo\ tables, one needs to keep in mind that the uniform-density torus is just an approximation of a non-uniform (clumpy) distribution of matter around the SMBH. This idea is illustrated in \autoref{fig:sketch}. We note that in cases where the torus cannot be approximated well with a smoothed distribution of clouds, such as when "holes" in the putative torus are much larger than individual clumps, our model may not be appropriate for constraining the covering factor. We believe our choice of geometry for \borustwo\ is a reasonable approximation based on the results in the literature; however, its assumptions will need to be tested with models assuming more complex geometries in the future. It is possible to define a covering factor (e.g., fraction of the sky covered with column density above some threshold, as seen from the SMBH at the center) and a typical column density (e.g., average over all obscured sightlines to the SMBH) for a wide variety of possible geometries. For any torus, the line-of-sight column density (\nhlos) can differ widely depending on its orientation with respect to the observer at a given time. Parameters \nhtor\ and \cftor\ therefore provide information on the material {\em outside} of our line of sight. 

In the paradigm described above, it is not difficult to understand how the tori in NGC\,7582 and IC\,5063 can simultaneously have a high \cftor\ and \nhtor\ in the CT regime, without CT absorption in the line of sight. NGC\,7582 may have a clumpy torus with \nh\,$\sim10^{25}$\,\cmmt\ clumps covering $\lesssim$20\,\% of the sky and the rest covered with \nh\,$\approx 5\times 10^{23}$\,\cmmt, which averages to $\approx3\times 10^{24}$\,\cmmt, in agreement with our modeling in \S\,\ref{sec:fitting}. This configuration can explain the previously observed CT state, as well as the average line-of-sight column density (see \citealt{rivers+2015-ngc7582} for a summary of previous X-ray observations of NGC\,7582). Our modeling also constrains the inclination so that $\cos$\,\thetainc\,$\approx$\,\cftor\,$\approx0.9$, implying that we are viewing at the torus close to its edge. In the uniform torus model, reprocessed emission from the inner side of the torus can only be observed for $\cos$\,\thetainc\,$>$\,\cftor, but a clumpy torus would have such lines of sight even for $\cos$\,\thetainc\,$<$\,\cftor\ (see \autoref{fig:sketch} and \autoref{fig:modcomp}). Constraints on \thetainc\ from fitting \borustwo\ should be interpreted in relation to \cftor, rather than in absolute terms. 

The \nustar\ data robustly exclude the possibility that the torus in 3C\,390.3 has a high covering factor; with \cftor\,$\lesssim 0.3$, its reprocessed component may simply be due to the outer part of the accretion disk. The NGC\,7582 torus is unlikely to be ring-like (\cftor\,$\simeq0.1$) or even disk-like (\cftor\,$\simeq0.5$). In \S\,\ref{sec:fitting-extra-nhtest} we presented fitting results for NGC\,2110 and IC\,5063 under the assumption that \nhtor\,$=$\,\nhlos, which yields acceptable, though not preferred, models for their \nustar\ spectra. With this assumption, the NGC\,2110 torus appears to be sphere-like but has two orders of magnitude lower average column density than the CT torus of NGC\,7582. The torus in IC\,5063 fits in between the other three, with its likely high covering factor and borderline CT average column density. This is already a step forward in testing the common assumption that all Seyfert-like AGN possess essentially the same kind of a torus.

We stress that the constraints presented in this paper are based on \nustar\ data alone, and can therefore be significantly improved in more detailed studies in the future. The data used for demonstration in this study are representative of a long \nustar\ observation in the case of 3C\,390.3 (100\,ks), a short snapshot observation of a very bright AGN in the case of NGC\,2110 ($\approx$20\,ks, but with photon statistics typical of a long exposure on an typical local Seyfert), and short observations of IC\,5063 and NGC\,7582, characteristic of the \nustar\ snapshot survey of the \swiftbat-selected AGN (B18). Inclusion of good-quality soft X-ray data, as well as self-consistently modeled additional epochs, can help constrain the torus parameters even further (see \citealt{yaqoob+2015-mrk3} and \citealt{guainazzi+2016-mrk3}, for the case of Mrk\,3 without and with \nustar\ data, respectively). We anticipate that many such studies will be done within the operational lifetime of \nustar.

\subsection{Implications for Previous Results Based\\on Phenomenological Models}
\label{sec:discussion-pexrav}

Despite the availability of empirically motivated torus models in recent years, a large fraction of the literature, and especially studies of large AGN samples, made use of disk reprocessing models such as \pexrav\ to approximate the torus contribution to AGN spectra. The spectral fitting examples presented in \S\,\ref{sec:fitting} already suggest a natural explanation for the very low Compton hump strengths measured using \pexrav\ in some AGN (notably, radio galaxies; \citealt{ballantyne-2007}, \citealt{tazaki+2011}). Very strong non-relativistic reprocessing signatures have been observed both in stacked hard X-ray data (\citealt{malizia+2003}, \citealt{ricci+2011}, \citealt{esposito+walter-2016}) and in spectral analyses of particular AGN (e.g., \citealt{rivers+2013}, \citealt{vasudevan+2013}, B18). Within our model, scaling of the Compton hump amplitude and shape corresponds to scaling of the torus covering factor.

The moderate strength of the narrow iron lines and the absence of a correspondingly strong Compton hump at the same time can be explained simply as a ring-like torus with CT column density and a low covering factor, such as in the case of 3C\,390.3. The Compton hump in that case is weaker and broader than in the \pexrav\ model, leading to a low normalization of the reprocessed component, as often found in the literature (e.g., \citealt{sambruna+2009-3c390d3}, \citealt{ballantyne+2014-3c382}, \citealt{king+2017}). Conversely, AGN with strongly peaked Compton humps may feature sphere-like tori with high covering factors, as demonstrated by NGC\,7582. However, the shape and strength of the Compton hump also scale with the torus column density, making the correspondence non-trivial in general.

In the lower panel of \autoref{fig:cf_indicators} we show an indication of a correlation between relative normalization of the \pexrav\ continuum (\rpex; taken from B18 and \citealt{lohfink+2015-3c390d3}) and the torus covering factor modeled in this work. In fact, the correlation may be stronger if all tori are assumed to have \nhtor\,$=10^{24}$\,\cmmt\ (fixed), which effectively makes \cftor\ account for part of the spectral diversity that would otherwise be accounted for by differences in \nhtor. Given the broad constraints for the non-CT torus in NGC\,2110 and the fact that other choices of fixed \nhtor\ weaken the observed trend, we refrain from further quantifying this relationship. A comprehensive comparison of covering factors and \rpex\ parameters for a large sample of \nustar-observed AGN will be presented in B18. We do stress that, in spite of the historical importance of models such as \pexrav, we expect that the community will make a point of moving away from these outdated models in favor of models such as \borustwo.

\begin{figure}
\begin{center}
\includegraphics[width=1.0\columnwidth]{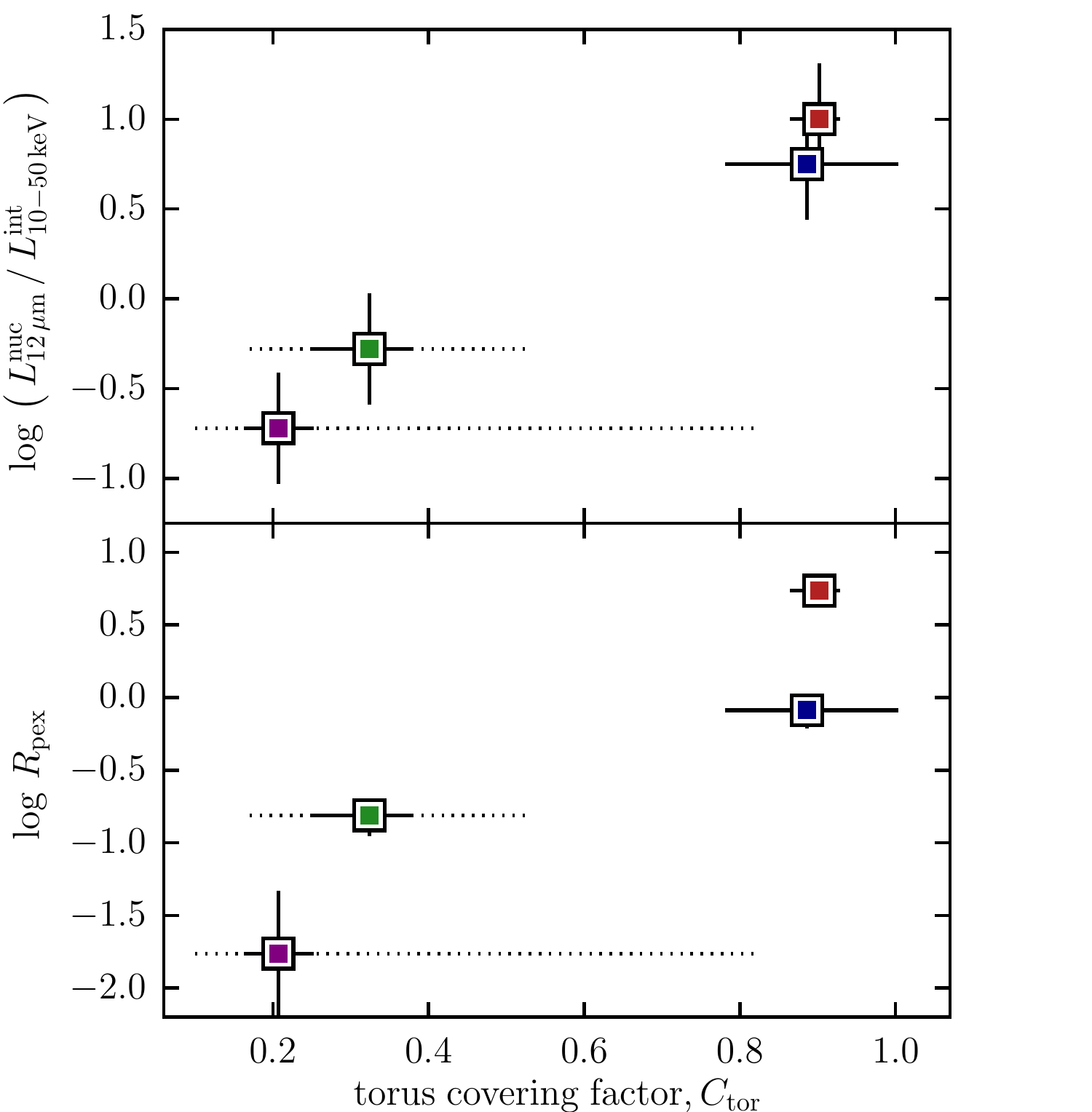}
\caption{Relationship between the fitted torus covering factor (\cftor) and two possible indicators of the covering factor. The upper panel shows the ratio of mid-infrared to X-ray luminosity, and the lower shows the relative normalization of the reprocessed continuum (\rpex\ parameter of the \pexrav\ model) from a phenomenological model fit to broadband X-ray spectrum. Marker colors correspond to different AGN, as in other figures. Marker edges and errorbars in plotted black correspond to \cftor\ constraints with best-fit torus column density (\nhtor) for each source. Dotted errorbars for 3C\,390.3 and NGC\,2110 illustrate how the uncertainty increases when \nhtor\ is left free to vary instead of being fixed at the best-fit value.
\label{fig:cf_indicators} }
\end{center}
\end{figure}

\vspace{0.5cm}
\subsection{Comparison with Constraints from Infrared Data}
\label{sec:discussion-infrared}

Through a simple energetics argument, the covering factor of the dusty torus can be related to the ratio of reprocessed, infrared luminosity to the intrinsic UV or X-ray luminosity (as a proxy of the bolometric output). Naively, a larger covering factor results in more intrinsic luminosity being intercepted, absorbed and re-radiated by the torus in the infrared. In the top panel of \autoref{fig:cf_indicators}, we use high spatial resolution 12-$\mu$m photometry from \citet{asmus+2014} and 10--50\,keV intrinsic luminosities based on \nustar\ data (measured from our best-fit models with removed absorption and reprocessed continuum) to show a possible link between the luminosity ratio and the covering factor. The trend is encouraging, and calls for further investigation with larger samples (e.g., Lanz et al., submitted). Recent calculations by \citet{stalevski+2016} suggest that this simple ratio may be effectively used as an indicator of the torus covering factor (e.g., in large surveys; \citealt{maiolino+2007}, \citealt{treister+2008}), provided that anisotropy of the disk and torus radiation is properly accounted for. 

Models for fitting torus SEDs in the infrared band have been available for a long time (see, e.g., \citealt{netzer-2015} for a recent review). These models have been used extensively for constraining torus properties in bright local AGN, as well as higher-redshift sources (e.g., \citealt{efstathiou+2013}, \citealt{roseboom+2013}, \citealt{podigachoski+2016}), to the extent possible with limited unresolved photometry. Directly comparing our \cftor\ constraints to the results from \citet{alonsoHerrero+2011}, we find that they are broadly consistent for all three AGN included in both studies, NGC\,2110, IC\,5063, and NGC\,7582. In all three cases, infrared-derived covering factors are high, $80-95\,\%$. Our \cftor\,$>0.8$ constraint for IC\,5063 is entirely consistent with this, as is \cftor\,$\approx 0.9$ for NGC\,7582. The covering factor for NGC\,2110 torus based on infrared data is nearly 100\,\%, which is an acceptable solution for the X-ray data, although the \nustar\ spectra indicate a preference for a lower value. Near-complete covering is obtained under the assumption that \nhtor\,$=$\,\nhlos\ for both NGC\,2110 and IC\,5063. While \citet{ichikawa+2015} also find a high dust covering factor for NGC\,2110 ($\approx90\,\%$) from infrared SED modeling, \citet{lira+2013} find a significantly lower dust covering factor ($\approx50\,\%$) for NGC\,7582, in disagreement with \citet{alonsoHerrero+2011} and our apparently very tight X-ray constraint. Torus column densities are very different between these three AGN, and may be indicative of a different spatial scale or nature of the reprocessing material (i.e., compact, possibly clumpy, torus versus gas and dust in the host galaxy).

Despite the fact that most infrared SED models include a geometrical measure such as the torus opening angle or, equivalently, covering factor, it is unclear to what extent this is comparable to \cftor\ of the X-ray torus. Dust-free gas, which affects X-ray reprocessing while not contributing to thermal dust emission, is thought to exist in the innermost regions of typical Seyferts (e.g., in the broad-line region; \citealt{gaskell+1981}). Comparing X-ray and infrared size estimates, \citet{gandhi+2015} have found that the bulk of fluorescent emission of iron likely originates from within the dust sublimation radius. The dust/infrared and gas/X-ray covering factors may therefore naturally differ depending on the overall geometry of the SMBH surroundings. Our model will enable some of the first systematic comparisons between the gas covering factor from the X-ray band and the constraints on dusty torus geometry derived from infrared modeling for single sources with high-quality data on the one hand, and large samples with lower-quality data on the other.

\subsection{Multi-wavelength Synergy in Future Studies}
\label{sec:discussion-future}

Thus far, only a small number of studies in the literature combine multi-wavelength probes of torus parameters with geometric constraints from high-quality hard X-ray spectroscopy; e.g., \citet{bauer+2015-ngc1068}, \citet{koss+2015-ngc3393}, \citet{masini+2016}. In a recent study of the obscured quasar IRAS\,09104$+$4109, \citet{farrah+2016-iras09104} combined X-ray, optical and infrared data in order to construct a self-consistent picture of its torus. While the short \nustar\ observation (15\,ks) did not provide constraints as tight as those derived from the infrared data, it is encouraging that both spectral bands independently yield results in agreement with optical (spectroscopic and polarimetric) and radio data. It would certainly be better to have an internally self-consistent model including both gas and dust distributions for spectra in both the infrared and X-ray bands; however, no such models have been published yet. In future work, \borus\ will be used to construct grids of spectral templates that enable simultaneous fitting of both infrared and X-ray data, including those with high energy resolution from an instrument similar to {\em Hitomi}/SXS \citep{mitsuda+2014}, such as {\em Athena}/XIFU \citep{barret+2016}, and {\em XARM}.

Placing constraints on the geometrical and physical torus parameters for single objects is possible from optical, infrared and radio observations. Ionization cone opening angles can be constrained from optical observations with the high spatial resolution of the {\em HST} (e.g., \citealt{schmitt+2003}, \citealt{fischer+2013}), and may be expected to correlate to some degree with the torus opening angle. For IC\,5063, \citet{schmitt+2003} found that the ionization cone has a half-opening angle of $\simeq30^{\circ}$, and that it is aligned well with the jet observed at radio frequencies. The broad-line region of IC\,5063 has been observed in polarized light \citep{inglis+1993}, indicating that it is present but hidden by intervening extinction. The torus geometry favored by our spectral modeling (\thetator\,$<40^{\circ}$, \thetainc\,$>50^{\circ}$, \lognhtorcmmt\,$\approx 23.9$) is remarkably consistent with these completely independent constraints.

Additional constraints from resolved ionization cone observations may be expected in the near future from {\em Chandra} and {\em JWST} (e.g., \citealt{maksym+2016-ngc3393}). Infrared photometry and spectroscopy at high spatial resolution \citep{ichikawa+2015}, interferometry \citep{burtscher+2013} and polarimetry  \citep{lopezRodriguez+2015-ngc1068} have significantly contributed to recent advances in probing AGN structures. Molecular gas observations resolving the torus scales in nearby AGN with ALMA are able to probe torus kinematics \citep{garciaBurillo+2016-ngc1068}. Some radio observations directly measure the orientation of the AGN structures with respect to the observer (e.g., jet, megamaser disk), while others are more indirect and model-dependent \citep{marin-2016}. Compared to these more traditional probes, constraints from the X-ray band have thus far been poorly explored, but they show promise for unique new insights into the nature of the AGN torus in the near future.

\section{Summary}
\label{sec:summary}

With the recent improvement in hard X-ray data quality brought about by \nustar, and the flexible empirically motivated spectral models, measuring the torus covering factor from the X-ray band is now possible for large samples of AGN. In this paper we present a new set of parametrized spectral templates, named \borustwo, made available to the public in the form of an \xspec\ table model, in order to facilitate studies of the torus geometry through X-ray spectroscopy. In calculation of the model spectra we assumed an approximately toroidal geometry with conical polar cutouts, following the popular \bntorus\ model of \citet{brightman+nandra-2011a}. \borustwo\ is an updated, expanded, and more flexible torus model that supersedes \bntorus.

Because \borustwo\ represents only the reprocessed spectral component (separated from the line-of-sight component), while featuring both the average column density and the covering factor as free parameters, it is applicable to a wide variety of AGN. In order to highlight its capabilities, we presented its application on four AGN observed with \nustar. These four examples cover different parts of the parameter space spanned by the column density ($22<$\,\lognhtorcmmt\,$<25.5$) and the covering factor ($0.1<$\,\cftor\,$<1.0$). Furthermore, we demonstrated how inclusion of multi-epoch data, external constraints and various assumptions can help with evaluating  or alleviating some systematic uncertainties. 

Finally, we compared our constraints on the torus covering factor with dust covering factors derived from modeling of infrared data, and found encouraging consistency. More detailed work will be required in order to understand the relationship between constraints from different wavelength regimes in terms of a physical interpretation. When combined self-consistently, the joint leverage of these different probes of torus geometry and orientation (not limited only to X-ray and infrared spectral modeling) should enable us to better characterize the complex geometry of the unresolvable innermost region surrounding SMBHs, and replace the proverbial donut-like AGN torus with a more realistic structure.

\acknowledgments{
The authors thank the anonymous referee for careful reading and constructive suggestions that improved the paper. The authors gratefully acknowledge financial support from NASA Headquarters under the NASA Earth and Space Science Fellowship Program, grant NNX14AQ07H (M.\,B.), the ASI/INAF grant I/037/12/0--011/13 (A.\,C.), the Caltech Kingsley visitor program (A.\,C.), FONDECYT 1141218 (C.\,R.), Basal-CATA PFB--06/2007 (C.\,R.) and the China-CONICYT fund (C.\,R.).

This work made use of data from the \nustar\ mission, a project led by the California Institute of Technology, managed by the Jet Propulsion Laboratory, and funded by the National Aeronautics and Space Administration. We thank the \nustar\ Operations, Software and Calibration teams for support with the execution and analysis of these observations. Furthermore, this research made use of the following resources: the NASA/IPAC Extragalactic Database (NED), operated by the Jet Propulsion Laboratory, California Institute of Technology, under contract with the National Aeronautics and Space Administration; the High Energy Astrophysics Science Archive Research Center Online Service, provided by the NASA/Goddard Space Flight Center; NASA's Astrophysics Data System; \texttt{matplotlib}, a Python library for publication quality graphics \citep{hunter+2007}.

\facility{\nustar}

}

\mbox{~}

\mbox{~}

\mbox{~}

\mbox{~}

\mbox{~}

\mbox{~}

\mbox{~}

\mbox{~}

\mbox{~}

\mbox{~}

\appendix

\section{Comparison with Other Torus Reprocessing Models}

In this Appendix we provide more details on the comparison of \borustwo\ with other publicly available models for X-ray reprocessing in the torus, namely \mytorus\ \citep{murphy+yaqoob-2009}, \etorus\ \citep{ikeda+2009}, and \ctorus\ \cite{liu+li-2014}. An overview of their main properties and parameters is presented in \autoref{tab:apptable}. A comparison with the \bntorus\ model \citep{brightman+nandra-2011a}, which shares the geometry chosen for \borustwo, but is less flexible and does not reproduce reprocessed spectra correctly, is given in \S\,\ref{sec:model-comparison}. Unless specified otherwise, for all comparisons we assume the intrinsic continuum to have $\Gamma=1.8$, and we select the highest available \ecut\ within the model wherever the option exists. As the abundance of iron can only be varied within the \borustwo\ model, we keep it fixed at the Solar value. 

We first compare a set of model reprocessed X-ray spectra with matched parameters in \autoref{fig:repcomp_new}: one set assuming a Compton-thin torus (\lognhtorcmmt\,$=23.5$, left column), and another assuming a Compton-thick torus (\lognhtorcmmt\,$=24.5$, right column). Both pole-on (\thetainc\,$=20^{\circ}$, solid lines) and edge-on (\thetainc\,$=84^{\circ}$, dashed lines) viewing angles are compared. Each of the panels shows the intrinsic continuum and its Thomson-scattered reflection (dotted lines) normalized to 0.3\,\% of the intrinsic continuum. This value of the relative normalization is chosen as an approximate lower end of the distribution for typical obscured AGN (e.g., \citealt{ricci+2017-bass}, B18), with only $\lesssim15$\,\% of obscured AGN showing lower contributions from such a component. Differences in the models much below this line are therefore of limited practical importance.

\begin{deluxetable*}{lcccc}
\tabletypesize{\small}
\tablewidth{0cm}
\tablecolumns{5}

\tablecaption{ Comparison of Parameters of Publicly Available Torus Reprocessing Models \label{tab:apptable} }

\tablehead{
  \colhead{} &
  \colhead{\borustwo} &
  \colhead{\mytorus} &
  \colhead{\etorus} &
  \colhead{\ctorus}
}

\startdata
reference & this work & \citet{murphy+yaqoob-2009} & \citet{ikeda+2009} & \citet{liu+li-2014} \\
\hline 
\multirow{2}{*}{geometry} & uniform-density sphere & \multirow{2}{*}{uniform-density torus} & similar to \borustwo, & same as \borustwo, \\
 & with polar cutouts &   & with a central cavity & but clumpy  \\
\hline 
\multirow{3}{*}{lines} & K$\alpha$: $A<31$ & K$\alpha$: Fe,\,Ni & \multirow{3}{*}{none} & K$\alpha$: Mg,\,Al,\,Si,\,S,\,Ar,\,Ca,\,Fe,\,Ni \\
 & K$\beta$: $A<31$ & K$\beta$: Fe &  & K$\beta$: Ca,\,Fe,\,Ni \\
 & Compton shoulder & Compton shoulder &  & no Compton shoulder  \\
\hline 
$\Gamma$ & 1.4--2.6 & 1.4--2.6 & 1.5--2.5 & 1.5--2.5 \\
\hline 
\multirow{2}{*}{\ecut\,/\,keV} & \multirow{2}{*}{20--2000} & 200\,(f)\,\tablenotemark{a} & 360 (f) for the $\Gamma$ range & \multirow{2}{*}{500\,(f)\,\tablenotemark{a}} \\
 &  & 500\,(f)\,\tablenotemark{a}  & 20--500 for $\Gamma=1.9$\,(f) & \\
\hline 
\cftor & 0.1--1.0 & 0.5\,(f) & 0.34--0.98 & 0.5\,(f) \\
\hline 
\thetainc\,/\,$^{\circ}$ & 19--87 & 0--90 & 1--89 & 19--87 \\
\hline 
\afe\,/\,\afesun & 0.1--10 & 1 (f) & 1 (f) & 1 (f) \\
\enddata
\tablecomments{\,(f) marks a fixed parameter. The information not found in the cited references may be found on the websites hosting the \xspec\ tables: \url{http://mytorus.com} (\mytorus), \url{https://heasarc.gsfc.nasa.gov/xanadu/xspec/models/etorus.html} (\etorus), \url{https://heasarc.gsfc.nasa.gov/xanadu/xspec/models/Ctorus.html} (\ctorus). }
\tablenotetext{a}{This is a sharp cut-off at the given energy instead of an exponential roll-over with a given scale.}

\end{deluxetable*}

Our choice of geometry for \borustwo\ is very similar to that of \citet{ikeda+2009}. Their publicly available model, \etorus, does not include fluorescent line emission, but the reprocessed continuum shape and normalization matches that of \borustwo\ remarkably well. A comparison is shown in \autoref{fig:repcomp_new} for three different torus opening angles corresponding to \cftor\,$=\,0.35$, 0.50, and 0.80. Due to different assumed geometry, a direct comparison with \mytorus\ and \ctorus\ models is less straightforward. In both cases, we can only make an approximate comparison for \cftor\,$=0.50$, which is fixed in those models. For both models, the average torus column density is related to the equatorial column density (which is a fitting parameter) via a factor that depends on geometry: $\pi/4$ for \mytorus\ (calculated exactly; \citealt{murphy+yaqoob-2009}) and 0.66 for \ctorus\ (estimated; \citealt{liu+li-2014}). Since the latter is a clumpy torus model, we show the spectra with the maximum number of clouds ($N_{\mbox{\scriptsize clo}}=10$) in order to approximate a uniformly filled torus as much as possible. Minor differences that can be attributed to geometry (e.g., possible {\em silver lining} effects from low-\nh\ regions close to the torus rim) are apparent from \autoref{fig:repcomp_new}, though for the most part, we find qualitative agreement between all models.

\begin{figure*}
\begin{center}
\includegraphics[width=0.85\textwidth]{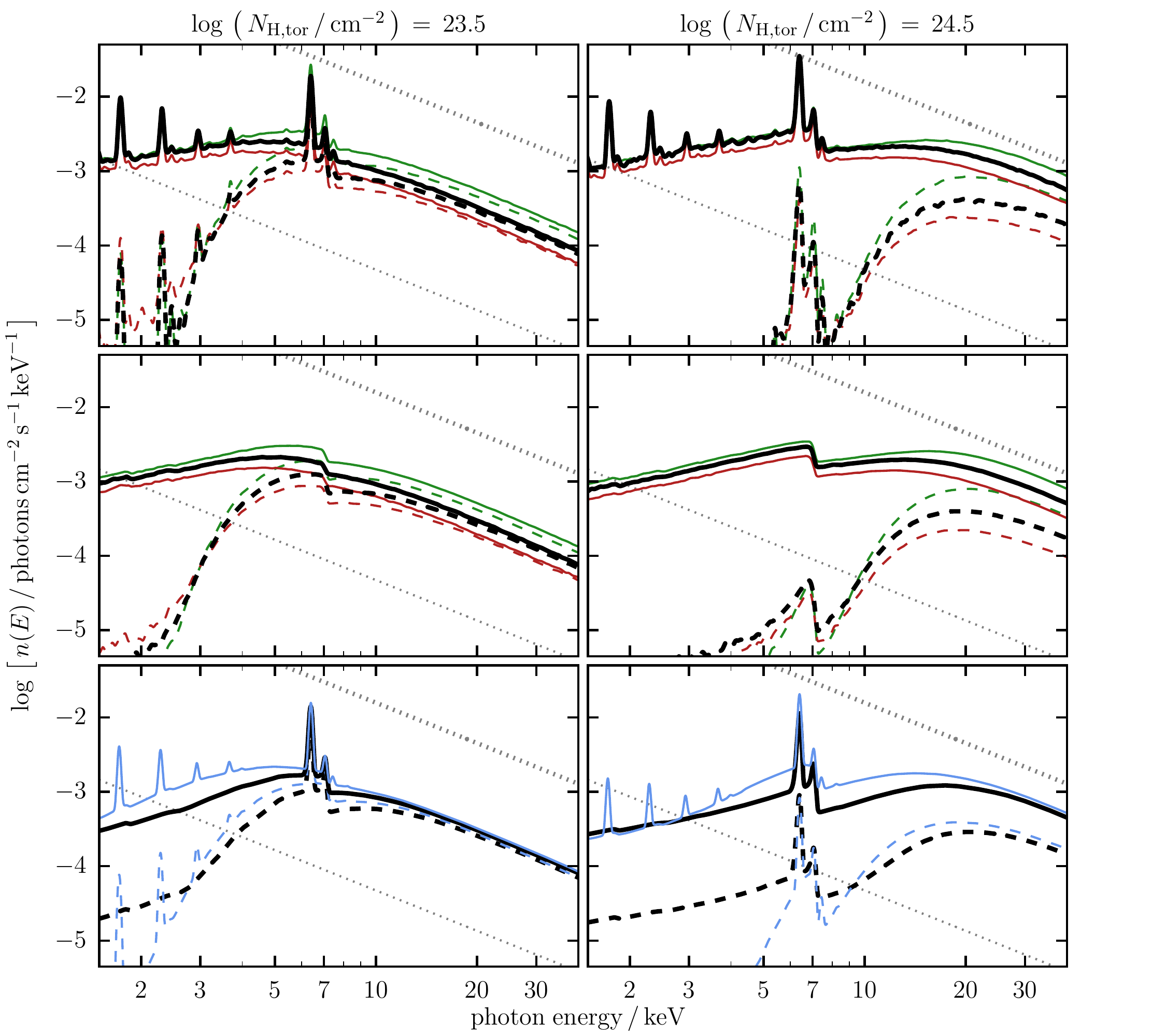}
\caption{Comparison of reprocessed spectra between \borustwo\ and existing torus-reprocessing models. The left column of panels shows reprocessed spectra for tori with average column density which is Compton-thin, and the right column shows spectra for Compton-thick tori. Dotted grey lines in each panel show the intrinsic continuum (upper line) and its Thomson-scattered reflection (lower line) at 0.3\,\% of the intrinsic flux. Solid lines show spectra for nearly pole-on inclination (\thetainc\,$=\,20^{\circ}$), while dashed lines show spectra for nearly edge-on inclination (\thetainc\,$=\,84^{\circ}$). The top row shows spectra from the \borustwo\ model presented in this paper, with covering factors 0.35 (red), 0.50 (black) and 0.80 (green). The middle row shows the same as the top row for the \etorus\ model \citep{ikeda+2009}. The bottom row shows the \mytorus\ \citep{murphy+yaqoob-2009} and \ctorus\ \citep{liu+li-2014} models with black and light blue lines, respectively. For the latter we assume the highest available cloud density ($N_{\mbox{\scriptsize clo}}=10$) in order to more closely approximate a uniform density assumed in the rest of the models. \label{fig:repcomp_new} }
\end{center}
\end{figure*}

In \autoref{fig:feew} we show a set of calculations of the equivalent width (EW) of the Fe\,K$\alpha$ fluorescent line at 6.4\,keV. In this calculation we assume that \nhlos\ is equal to \nhtor\ for any viewing angle that intersects the torus and zero otherwise. Dependence of \feew\ is given for four different viewing angles: edge-on (\cosi\,$=0.05$), pole-on (\cosi\,$=0.95$), and below and above the torus rim. The latter two depend on the opening angle; we use angles with $\Delta$\,\cosi\,$=\pm\,0.05$ around \cftor\,$=0.25$, 0.50, and 0.75. For \cftor\,$=0.50$ (second panel from the left in \autoref{fig:feew}), these curves can be directly compared with, e.g., Figure~8 in \citet{murphy+yaqoob-2009}, and Figure~6 in \citet{furui+2016}. Curves with smaller and larger covering factor may be compared to, e.g., Figure~3 in \citet{ghisellini+1994}, and Figure~13 in \citet{ikeda+2009} (also Figure~3 in \citealt{brightman+nandra-2011a}, keeping in mind the issues discussed in \S\,\ref{sec:model-comparison}). Based on these comparisons, we conclude that fluorescent line emission in \borustwo\ is in agreement with previous work. 

A feature that is perhaps unexpected, given that observations rarely yield \feew\ greater than a few keV (e.g., \citealt{dadina-2008}, \citealt{fukazawa+2011}, \citealt{boorman+2016-ic3639}), is that in some cases our curves extend up to $\sim100$\,keV. However, this is simply due to the fact that in this calculation EW is evaluated against the transmitted and the reprocessed continuum components, both of which are heavily absorbed for nearly edge-on inclination in the Compton-thick regime. In reality, a small amount of off-nuclear Thomson-scattered secondary continuum (or contributions from the host galaxy) is sufficient to limit \feew\ to $<5$\,keV. This is illustrated in the rightmost panel of \autoref{fig:feew}, where we compare edge-on curves computed without (dotted lines) and with (solid lines) a small contribution from the Thomson-scattered continuum normalized to 0.3\,\% of the intrinsic continuum.

\begin{figure*}
\begin{center}
\includegraphics[width=1.0\textwidth]{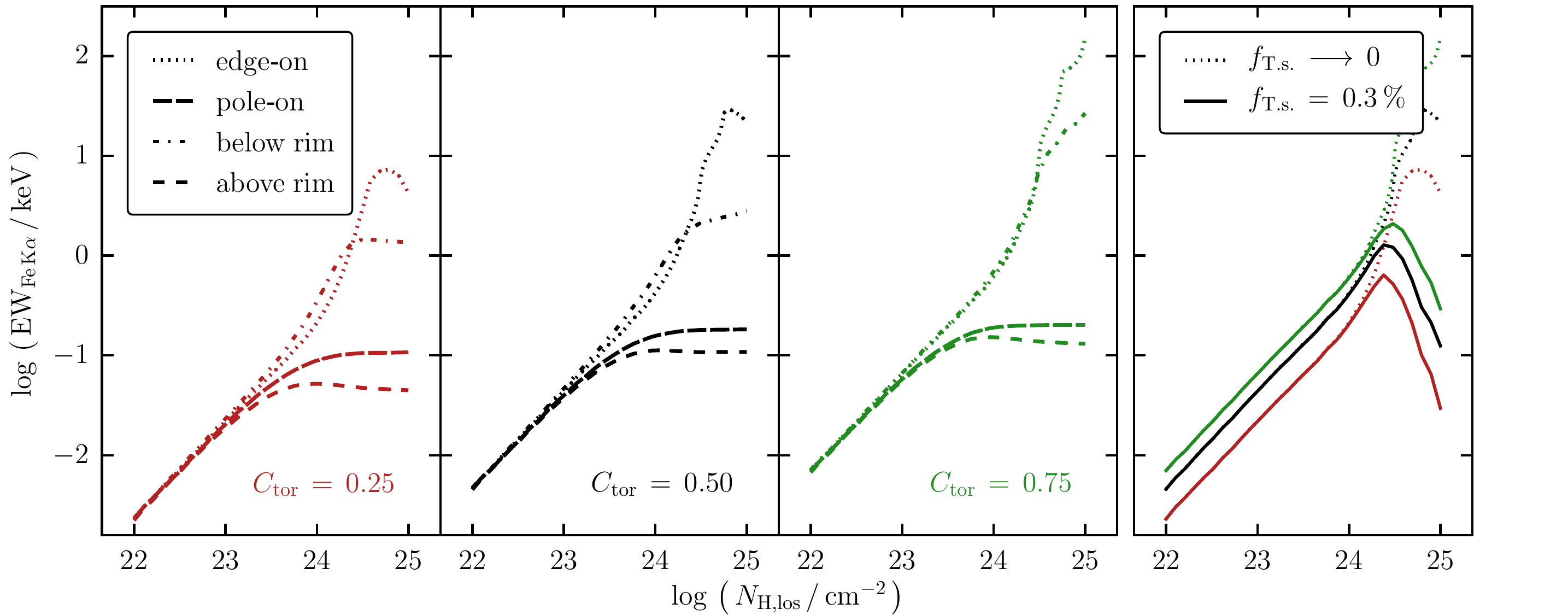}
\caption{ Equivalent width of the Fe\,K$\alpha$ line (\feew) as a function of \nhlos, which is assumed to be equal to \nhtor\ for lines of sight through the torus. The first three panels from the left show the run of \feew\ for four different viewing angles with respect to the rim of the torus, for tori with covering factors 0.25 (first), 0.50 (second) and 0.75 (third panel). The rightmost panel shows the effect of a small contribution of a Thomson-scattered component to the continuum, which limits the growth of \feew\ in the Compton-thick regime. \label{fig:feew} }
\end{center}
\end{figure*}

\end{document}